\documentclass[notoc,preprintnumbers]{JHEP3}
\usepackage{cite,color,psfrag,url,latexsym,graphicx,epstopdf,slashed,xspace,hyperref,enumitem,lineno}
\let\normalcolor\relax

\newcommand{\pt}{$p_{T}$}
\newcommand{\eqref}[1]{(\ref{#1})}
\newcommand{\text}[1]{{\rm #1}}
\newcommand{\jw}{\textsc{Jewel}~}

\DeclareRobustCommand{\Ref}[1]{Ref.~\cite{#1}}

\newenvironment{changemargin}[2]{%
  \begin{list}{}{%
    \setlength{\topsep}{0pt}%
    \setlength{\leftmargin}{#1}%
    \setlength{\rightmargin}{#2}%
    \setlength{\listparindent}{\parindent}%
    \setlength{\itemindent}{\parindent}%
    \setlength{\parsep}{\parskip}%
  }%
  \item[]}{\end{list}}

\newcommand{\be}{\begin{equation}}
\newcommand{\ee}{\end{equation}}
\newcommand{\cmb}{\begin{changemargin}}
\newcommand{\cme}{\end{changemargin}}
\newcommand{\bea}{\begin{eqnarray}}
\newcommand{\eea}{\end{eqnarray}}

\preprint{MIT-CTP 4947\\ WSU-HEP 1802}
\title{Probing heavy ion collisions using quark and gluon jet substructure}

\author{Yang-Ting Chien $^{a}$ and Raghav Kunnawalkam Elayavalli $^{b,c}$\\
$^{a}$ Center for Theoretical Physics\\
$~$ Massachusetts Institute of Technology, Cambridge, MA 02139\\
$^{b}$ Department of Physics and Astronomy\\
$~$ Wayne State University, Detroit, MI 48201\\
$^{c}$ Department of Physics and Astronomy\\
$~$ Rutgers, the State University of New Jersey, New Brunswick, NJ 08901
}

\abstract{
We study the phenomenon of jet quenching utilizing quark and gluon jet substructures as independent probes of heavy ion collisions. We exploit jet and subjet features to highlight differences between quark and gluon jets in vacuum and in a medium with the jet-quenching model implemented in \textsc{Jewel}. We begin with a physics-motivated, multivariate analysis of jet substructure observables including the jet mass, the radial moments, the $p_T^D$ and the pixel multiplicity. In comparison, we employ state-of-the-art image-recognition techniques by training a deep convolutional neutral network on jet images. To systematically extract jet substructure information, we introduce the telescoping deconstruction framework exploiting subjet kinematics at multiple angular scales. We draw connections to the soft-drop subjet distribution and illuminate medium-induced jet modifications using Lund diagrams. We find that the quark gluon discrimination performance worsens in heavy ion jets due to significant soft event activity affecting the soft jet substructure. Our work suggests a systematically improvable framework for studying modifications to quark and gluon jet substructures and facilitating direct comparisons between theoretical calculations, simulations and measurements in heavy ion collisions.
}

\begin{document}

\section{Introduction}
\label{sec:intro}

The jet quenching phenomenon observed in experiments at the Relativistic Heavy Ion Collider (RHIC) \cite{Adcox:2001jp,Adler:2002xw,Adcox:2004mh,Arsene:2004fa,Back:2004je,Adams:2005dq}
and the Large Hadron Collider (LHC)~\cite{Aamodt:2010jd,Aad:2010bu,Chatrchyan:2011sx,Aamodt:2011vg,CMS:2012aa,Chatrchyan:2012nia,Abelev:2012hxa,Chatrchyan:2012gt,Aad:2012vca,
Abelev:2013kqa,Aad:2013sla,Chatrchyan:2013exa,Aad:2014bxa,Khachatryan:2014bva,
Adam:2015ewa,Adam:2015doa,Aad:2015bsa,Khachatryan:2015lha,Aad:2015wga,
Khachatryan:2016odn,Khachatryan:2016xxp,Sirunyan:2016znt,Khachatryan:2016ypw,
Sirunyan:2017isk,Sirunyan:2017oug,Sirunyan:2017lzi,Sirunyan:2017jic,Sirunyan:2017xss,Sirunyan:2017qhf}
has since become an essential hard probe of the strongly interacting medium produced in heavy ion collisions (see \cite{Connors:2017ptx} for a review on heavy ion jet measurements). The dramatic suppression of hadron and jet cross sections is understood using the medium-induced energy loss picture \cite{Gyulassy:1993hr,Wang:1994fx,Zakharov:1996fv,Zakharov:1997uu,Baier:1996kr,Baier:1998kq,
Gyulassy:2000er,Gyulassy:2000fs,Wiedemann:2000za,Wang:2001ifa,
Arnold:2001ba,Arnold:2001ms,Arnold:2002ja,Casalderrey-Solana:2014bpa}. In order to quantitatively extract key features of jet-medium interactions and the underlying medium dynamics, a precise understanding of the redistribution of jet energy in heavy ion collisions is becoming essential. With the proliferation of accurate comparisons between theoretical calculations and measurements \cite{Chatrchyan:2013kwa,Khachatryan:2016tfj,Chatrchyan:2012gw,Aad:2014wha,Chatrchyan:2014ava,Aaboud:2017bzv,Sirunyan:2018qec,Khachatryan:2016erx,
Acharya:2017goa,Sirunyan:2017bsd,CMS:2017xdn}, the field of heavy ion jet physics has entered the era of precision jet substructure and jet cross section studies.

Jet substructure measurements provide concrete and consistent physics information of the modification of jets in a heavy ion environment. From the measurements of the jet shape \cite{Ellis:1992qq,Chatrchyan:2013kwa,Khachatryan:2016tfj,Seymour:1997kj,Li:2011hy,Li:2012bw,Vitev:2008rz,Vitev:2009rd} and jet fragmentation function \cite{Procura:2009vm,Chatrchyan:2012gw,Aad:2014wha,Chatrchyan:2014ava,Aaboud:2017bzv,Sirunyan:2018qec}, the community has recently established that both transverse and longitudinal momentum distributions inside jets are significantly modified. Since jet substructures depend strongly on the partonic origin of jets \cite{Gallicchio:2011xq,Gallicchio:2012ez,Chien:2012ur,Chien:2015ctp,Chien:2014nsa}, a change in the fraction of quark-initiated jets and gluon-initiated jets at the LHC may significantly contribute to substructure modifications observed in experiments. It has recently been realized that an increase of the quark jet fraction in heavy ion collisions can contribute to the narrowing of jet cores as suggested in various inclusive jet substructure measurements~\cite{Chien:2015hda,Spousta:2015fca}. With jet samples consisting of different quark and gluon jet fractions, for example comparing inclusive jet and photon-tagged jet measurements, one can disentangle the effect on jet substructure modifications due to the changes in quark and gluon jet fractions caused by jet-medium interactions. This then allows one to focus on studying how quark and gluon jets are differently modified and use them as independent jet quenching probes. Note that the above jet modification picture implies that, with the high-purity quark jets in photon-tagged jet samples, the narrowing effect of jet cores should decrease. On the other hand, certain jet quenching models predict a universal narrowing of jet cores \cite{KunnawalkamElayavalli:2017hxo,Milhano:2017nzm,Casalderrey-Solana:2016jvj,Brewer:2017fqy} and emphasize contributions from medium responses \cite{Tachibana:2017syd}. Selecting purer samples of quark and gluon jets will provide the opportunity for directly studying how the medium responds to different hard probes.

The use of quark jets and gluon jets as independent hard probes of heavy ion collisions is closely related to the tagging of these two probes, which is an outstanding problem \cite{Gras:2017jty,Frye:2017yrw}. The lowest-order feature that separates quark jets from gluon jets is the different color charges carried by the jet-initiating partons. The Casimir factors of quark jets and gluon jets are $C_F = 4/3$ and $C_A=3$, respectively, and the larger color charge of gluons is expected to result in broader spread of the radiation inside such jets. Additional distinctive features can be included to improve the tagging by combining a variety of jet substructure variables in a multivariate analysis, similar to how heavy flavor jets are identified at the LHC \cite{Chatrchyan:2012jua, Aad:2015ydr}. Driven by the need to identify and study boosted objects at the LHC, the high energy physics community has made significant advances on the use of machine-learning techniques leading to higher tagging efficiency in a wide kinematic range \cite{Komiske:2016rsd, Larkoski:2017jix}.

Jet quenching studies of the past decades have made compelling, qualitative strides by comparing jets in proton-proton and heavy ion collisions. In this paper, we systematically study the tagging of quark and gluon jets in both proton-proton and lead-lead collisions making full use of their substructure information. This indirect, complimentary approach studies the  differences in the modification of quark v.s. gluon jets, thus disentangling modifications of their common features in a multivariate analysis. By first comparing probes in the same collision system, it can also help reduce the systematic uncertainty in experimental measurements. We will demonstrate our general method using Monte Carlo simulations and take \jw as a concrete example \cite{Zapp:2013zya,KunnawalkamElayavalli:2016ttl}. However, the method is easily applicable to other heavy ion jet simulations \cite{Armesto:2009fj,Casalderrey-Solana:2016jvj,Cao:2017zih} and experimental data as well.

We will use three classes of methods and gain insights from comparing their performances. We combine in a multivariate analysis, five physics-motivated and representative jet substructure variables which are effective in quark gluon discrimination \cite{Gallicchio:2012ez}. They include the jet mass, two radial moments including the girth, the $p_T^D$ and the pixel multiplicity. These variables are combined and studied in a multi-layer perceptron (MLP) with two hidden layers to extract correlations among the variables. Next, we employ a modern 2-dimensional image recognition technique involving a deep convolutional neural network (DCNN). The network is trained on discretized images of quark jets and gluon jets on pseudo rapidity-azimuthal angle $(\eta,\phi)$ plane which forms a fixed-dimensional representation of the jet energy distribution. Each jet image undergoes standard pre-processing as described in \cite{deOliveira:2015xxd}. Finally, in order to systematically extract the complete jet substructure information, we apply the newly developed telescoping deconstruction (TD) method. At each order $N$, TD deconstructs a jet into its fragmentation basis with regard to $N$ axes that are defined using the winner-take-all (WTA) recombination scheme \cite{Bertolini:2013iqa}. The corresponding subjet expansion can be truncated at any finite order thus allowing us to order-by-order examine the information carried by subleading subjets. These three methods are used in discriminating quark and gluon jets, as well as discriminating jets in proton-proton and heavy ion collisions.

Due of the huge underlying event background in heavy ion collisions, jets with smaller radii are more reliably reconstructed and theoretically well understood \cite{Dasgupta:2014yra,Chien:2015cka,Becher:2015hka,Kang:2016mcy}. In order to minimize the background contamination within the experimentally allowed angular resolution, we study jets reconstructed using the anti-$k_t$ algorithm \cite{Cacciari:2008gp} with $R=0.4$. Within the catchment area of such jets, the significant background contamination needs to be subtracted before meaningful jet observable measurements can be made. In order to study multiple jet substructure observables and their correlations simultaneously, it necessarily requires constituent-level, observable-independent background subtraction methods \cite{Soyez:2012hv,Cacciari:2014gra,Berta:2014eza,Bertolini:2014bba,Komiske:2017ubm}. Furthermore, various jet grooming techniques \cite{Ellis:2009su,Ellis:2009me,Krohn:2009th,Dasgupta:2013ihk,Larkoski:2014wba} systematically remove soft radiation and highlight contributions from hard, collinear particles to jet observables. We will draw connections from telescoping deconstruction to the soft-drop groomed momentum sharing $z_g$ and groomed jet radius $r_g$ variables \cite{Larkoski:2014wba} and discuss qualitatively new jet substructure observables within the context of the telescoping deconstruction framework. During the course of the paper, we also use Lund diagrams \cite{Andersson1989,Salam:2016yht} to represent and showcase significant regions of phase space in \jw in the context of the soft drop grooming procedure.

The rest of the paper is organized as follows. In Sec.~\ref{sec:sample} we describe the jet samples we generate using the \jw simulation. In Sec.~\ref{sec:qg} we give details about the multivariate analysis, the jet image, and the telescoping deconstruction methods, as well as the corresponding jet substructure distributions. We also use Lund diagrams to illustrate the regions of phase space in jet formation separated by the soft-drop procedure. In Sec.~\ref{sec:results} we compare the performances of the methods and discuss the physical interpretations. We conclude in Sec.~\ref{sec:conc} and raise awareness of such methods by giving an outlook to future studies.

\section{Quark and Gluon Jet Samples}
\label{sec:sample}

To use quark and gluon jets as independent probes of heavy ion collisions, we need to first define and identify the respective probes. In this paper, we define jet flavor based on the parton from hard scattering that initiates the parton shower evolution. We use the \jw Monte Carlo simulation, a perturbative jet evolution framework to generate proton-proton and central (0-20\%) lead-lead collisions at $\sqrt{s}=2.76$ TeV using the standard setup \cite{Zapp:2013zya}. We use the prompt photon production channels \cite{KunnawalkamElayavalli:2016ttl} $q +\gamma$ and $g +\gamma$ to generate quark and gluon jet samples, respectively, as an attempt to prepare two sets of jets with distinct quark and gluon jet fractions. It is possible to construct jet samples by performing hard process and kinematical selections (such as dijet v.s. $Z$+jet) to enhance either the gluon or the quark jet fraction. On the other hand, unsupervised or weakly supervised learning algorithms \cite{Dery:2017fap, Cohen:2017exh, Metodiev:2017vrx,Komiske:2018oaa,Metodiev:2018ftz} can also be employed in the near future on such samples in order to treat both data and simulations on equal footing.

In each simulated event, jets are reconstructed using the anti-$k_t$ algorithm~\cite{Cacciari:2008gp} as implemented in \textsc{FastJet}~\cite{Cacciari:2011ma} with jet radius $R = 0.4$. We impose the following kinematic cuts,
\begin{equation}
    p^{\gamma}_T > 100~{\rm GeV}, ~~~ |\eta^{\rm \gamma}|<1.5, ~~~ p^{\rm jet}_T > 50~{\rm GeV}, ~~~ |\eta^{\rm jet}|<1.5, ~~~\Delta \phi_{\gamma, {\rm jet}} > 2\pi/3\;,
\end{equation}
where $p_T$, $\eta$ and $\phi$ are the transverse momentum, pseudo rapidity and azimuthal angle, respectively, to select high $p_T$ jets recoiling against the prompt photon. Jet quenching in central PbPb collisions in \jw is calculated based on a medium model consisting of thermal scattering centers undergoing Bjorken expansion. The medium formation time is set to $\tau_\text{i}=0.6 $ fm along with an initial temperature $T_\text{i}=485$ MeV based on hydrodynamic calculations~\cite{Shen:2012vn,Shen:2014vra}. We use the \textsc{CTEQ6LL} \cite{Pumplin:2002vw} and the \textsc{EPS09}~\cite{Eskola:2009uj} parton distribution functions in \textsc{LHAPDF-5}~\cite{Whalley:2005nh}. We include recoil partons in our PbPb simulations that originate from elastic and inelastic scattering with thermal scattering centers due to interactions with propagating hard scattered partons. It was recently shown that the recoil contributions are important in \jw to accurately describe several qualitative features as seen in jet substructure modifications measured at the LHC \cite{KunnawalkamElayavalli:2017hxo,Milhano:2017nzm}. In order to remove the thermal component of the recoil partons, we adopt the background subtraction procedure introduced in \cite{KunnawalkamElayavalli:2017hxo,Milhano:2017nzm}. The \jw simulation produces HepMC~\cite{Dobbs:2001ck} files which are processed using a \textsc{Rivet} analysis framework~\cite{Buckley:2010ar}.

To emulate finite detector effects on the jet angular resolution, we discretize jets using $\eta-\phi$ grids of size $0.08 \times 0.08$ respecting the current angular resolution of the CMS experiment's inherent hadronic calorimeter resolution at the LHC. To study how such discretization as well as background subtraction affect the analysis, we use the following three different sets of jets in heavy ion collisions, all with the medium recoil contributions,
\begin{itemize}
	\item neither discretization nor background subtraction
	\item with discretization but without background subtraction
	\item with discretization and the GridSub \cite{KunnawalkamElayavalli:2017hxo} background subtraction.
\end{itemize}
For jets in proton-proton collisions, we use the sets of jet samples with and without discretization. In experiments, the fluctuating, underlying event background is subtracted via specific background subtraction techniques. Since \jw does not generate full underlying events in quenched samples, we can not directly apply those background subtraction methods. With the application of the grid subtraction (GridSub) technique, the collection of jets in \jw represent ideally subtracted jets in the experiment. This enables us to focus on and highlight the effects of jet quenching, as opposed to additional, nontrivial smearing effects.

\section{Quark and Gluon Jet Substructure}
\label{sec:qg}
\begin{figure}[t]
	   \centering
	   \includegraphics[width=0.7\textwidth]{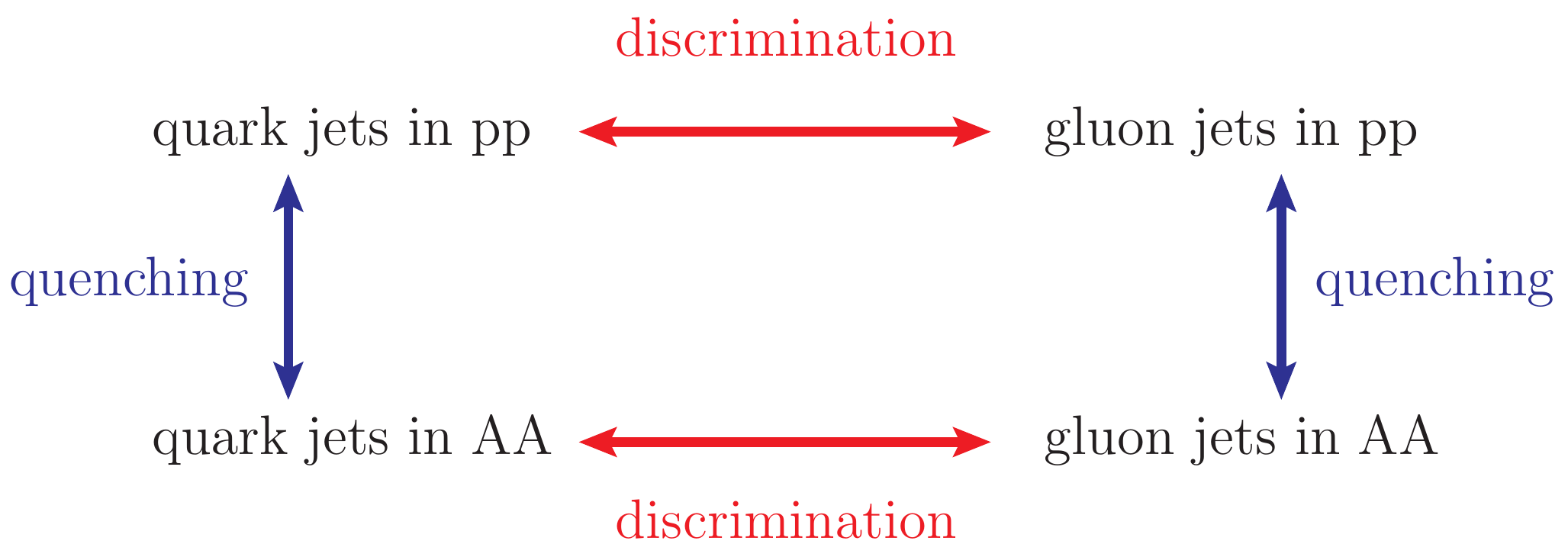}
	   \caption{Illustration of the interplay between jet quenching (vertical) and quark gluon discrimination (horizontal). }
	   \label{fig:quenching_discrimination}
\end{figure}

We use quark jets and gluon jets as different probes of heavy ion collisions and compare their modification patterns. In doing so one attempts to identify qualitatively different jet features that are sensitive to aspects of quark and gluon quenching mechanisms. Such direct approaches are closely related to the discrimination of quark and gluon jets, where the goal is to exploit jet features which emphasize differences between quark jets and gluon jets. By comparing quark and gluon jet substructures in the same collision system, one can quantify the respective jet modification. This method also helps to disentangle jet quenching effects on common features between quark and gluon jets, most notably the jet transverse momentum, and cancel systematic uncertainties in experiments. A schematic illustration of possible ways of exploiting quark and gluon jet substructure is shown in Figure~\ref{fig:quenching_discrimination} where the interplay between quenching (vertical arrows) and the jet flavor difference (horizontal arrows) are highlighted.

Every jet observable is defined as a function of the constituent momenta and angles from jet axes so as to probe the entire jet substructure with a single number. One can choose a set of jet observables to be a representation of jets, and we discus several approaches which faithfully encode the jet information.

\subsection{Physics-motivated Multivariate Analysis}
\label{sec:mva}

\begin{figure}[h]
	   \centering
	   \includegraphics[width=0.48\textwidth]{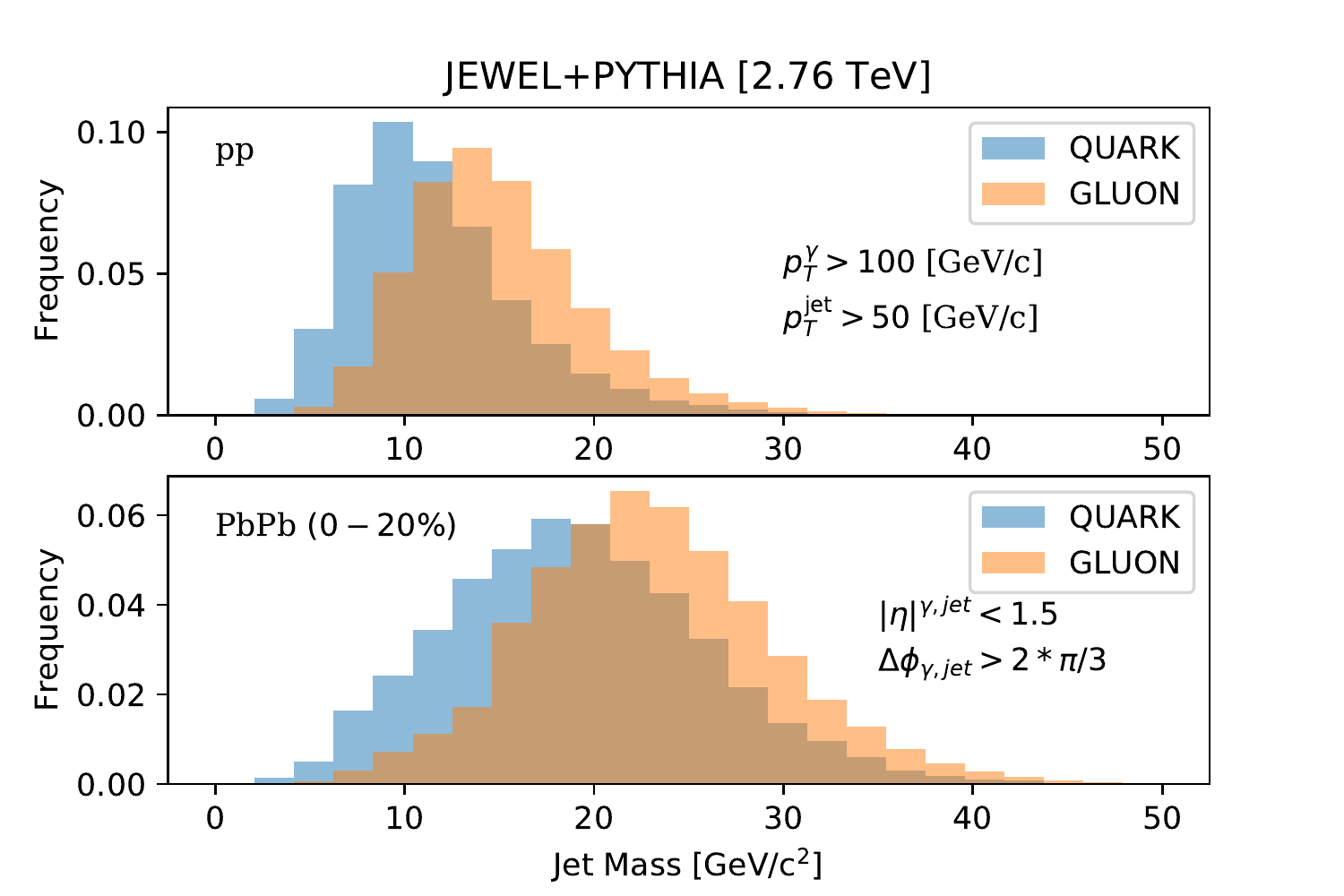}
	   \includegraphics[width=0.48\textwidth]{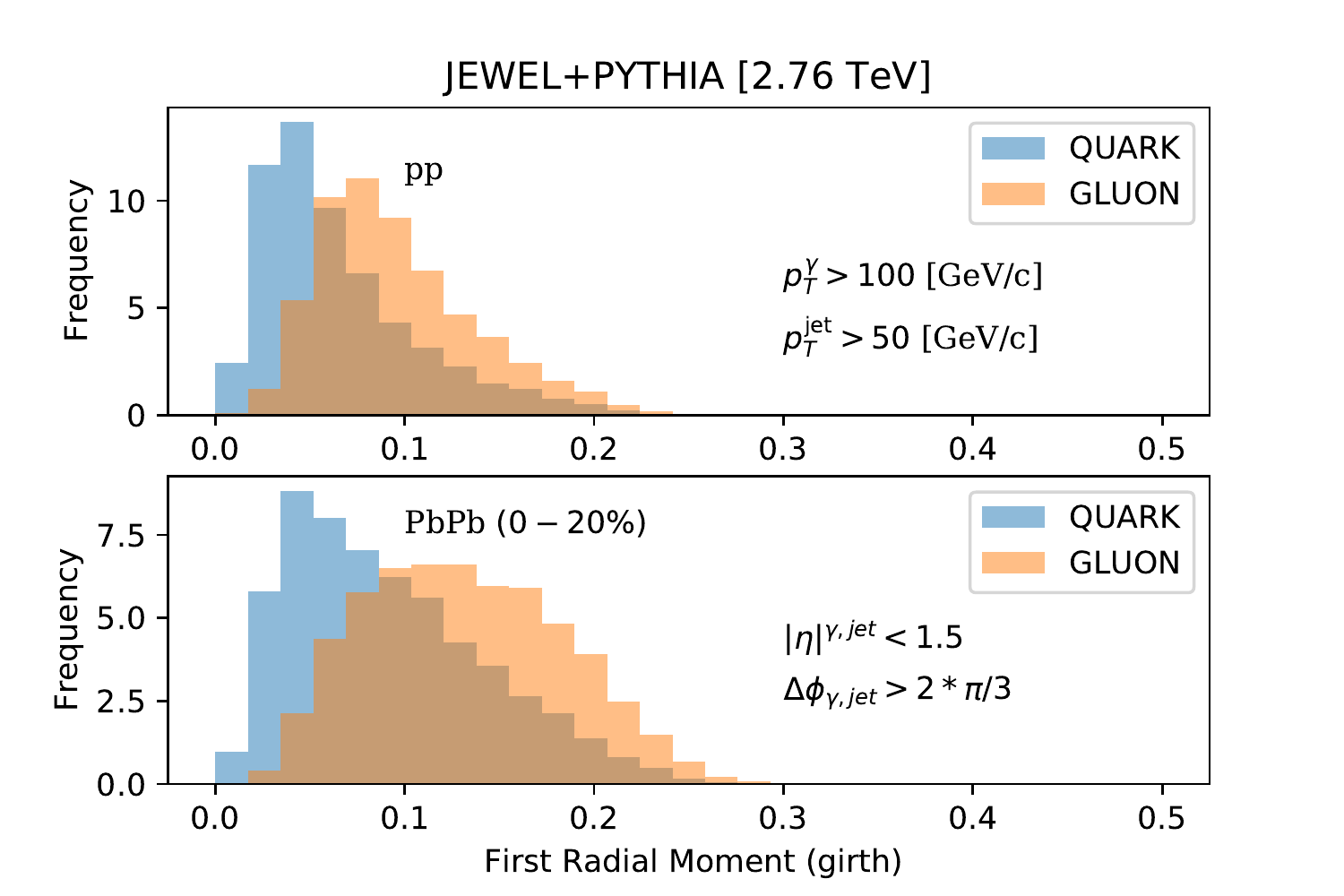}
	   \includegraphics[width=0.48\textwidth]{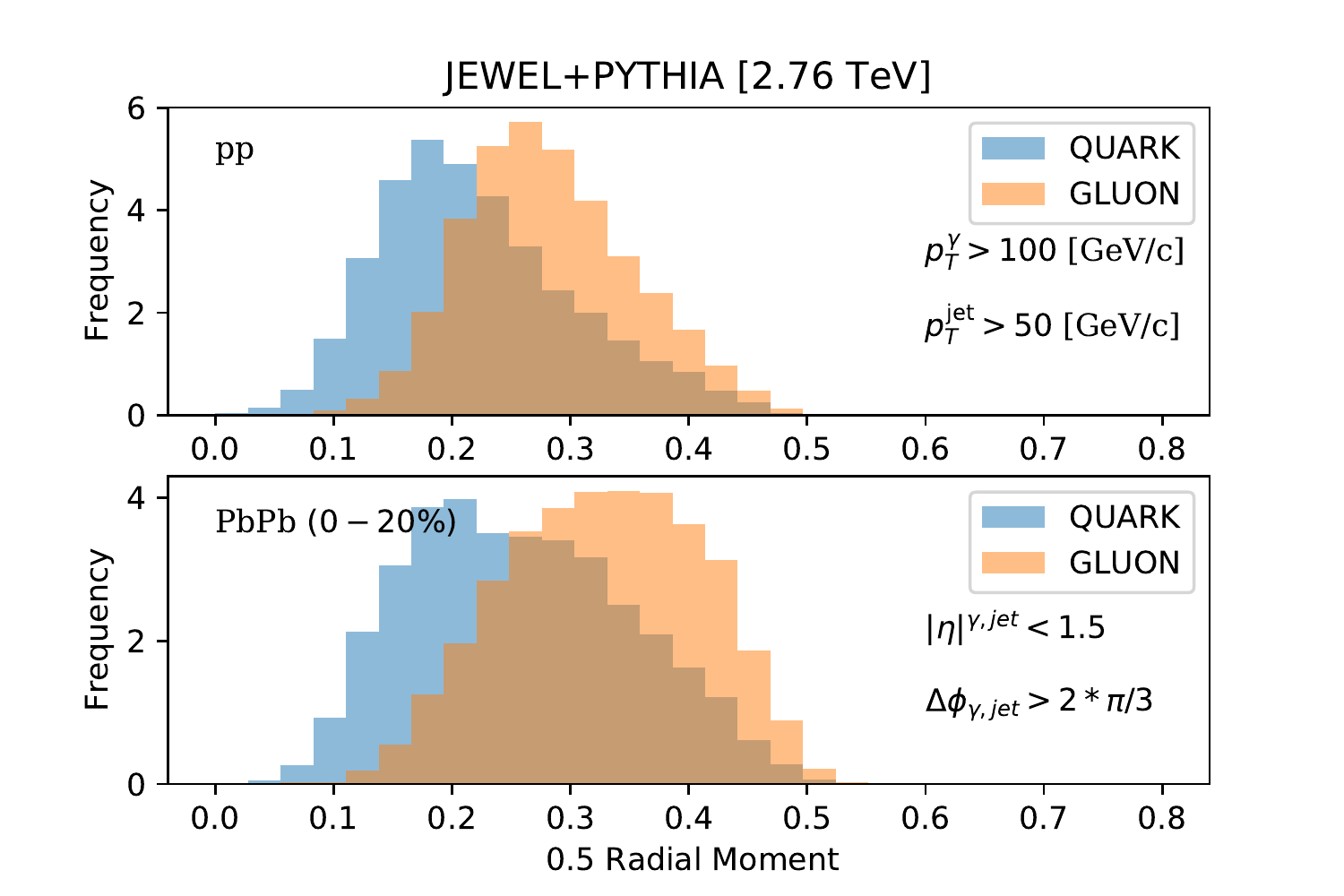}
	   \includegraphics[width=0.48\textwidth]{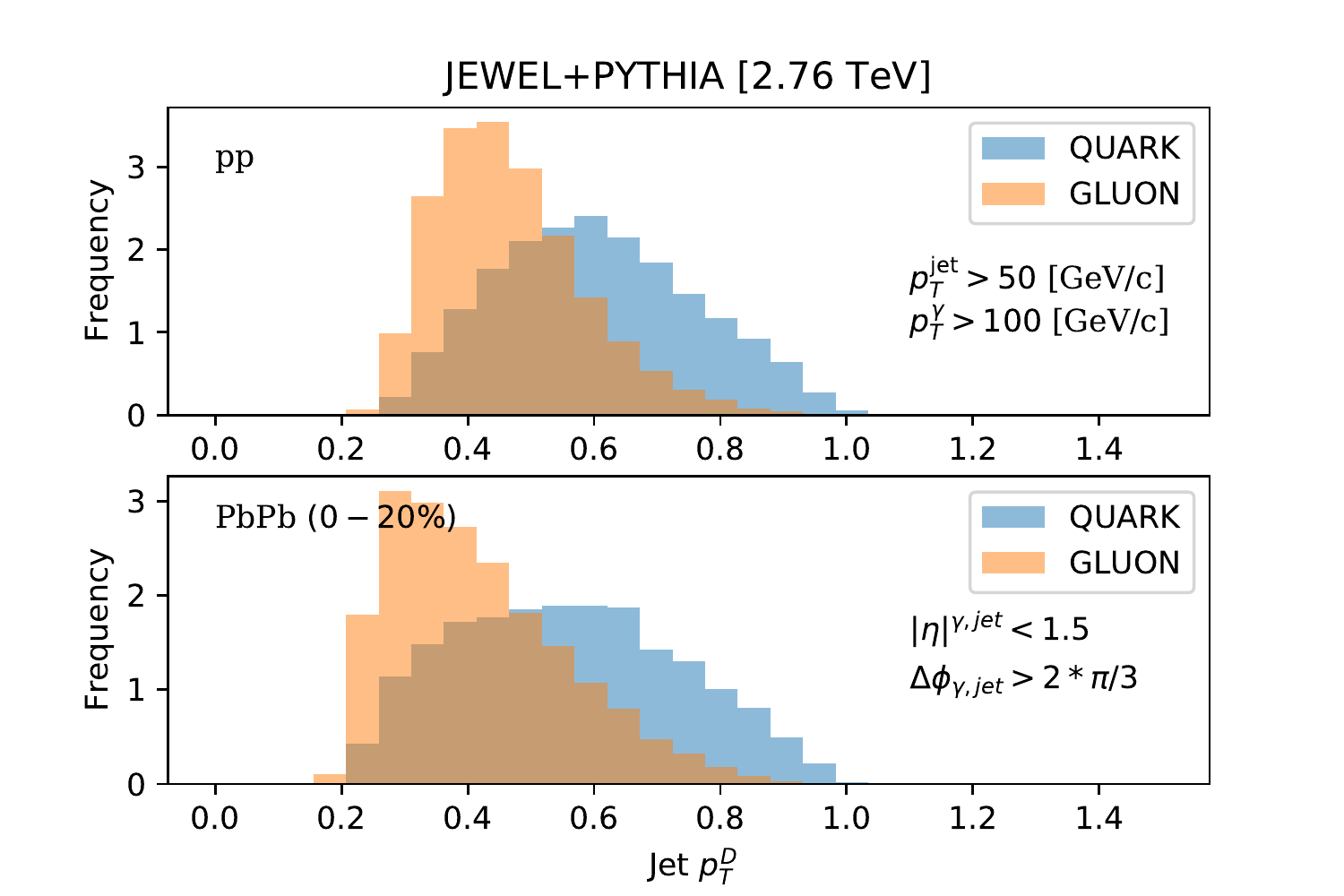}
	   \includegraphics[width=0.48\textwidth]{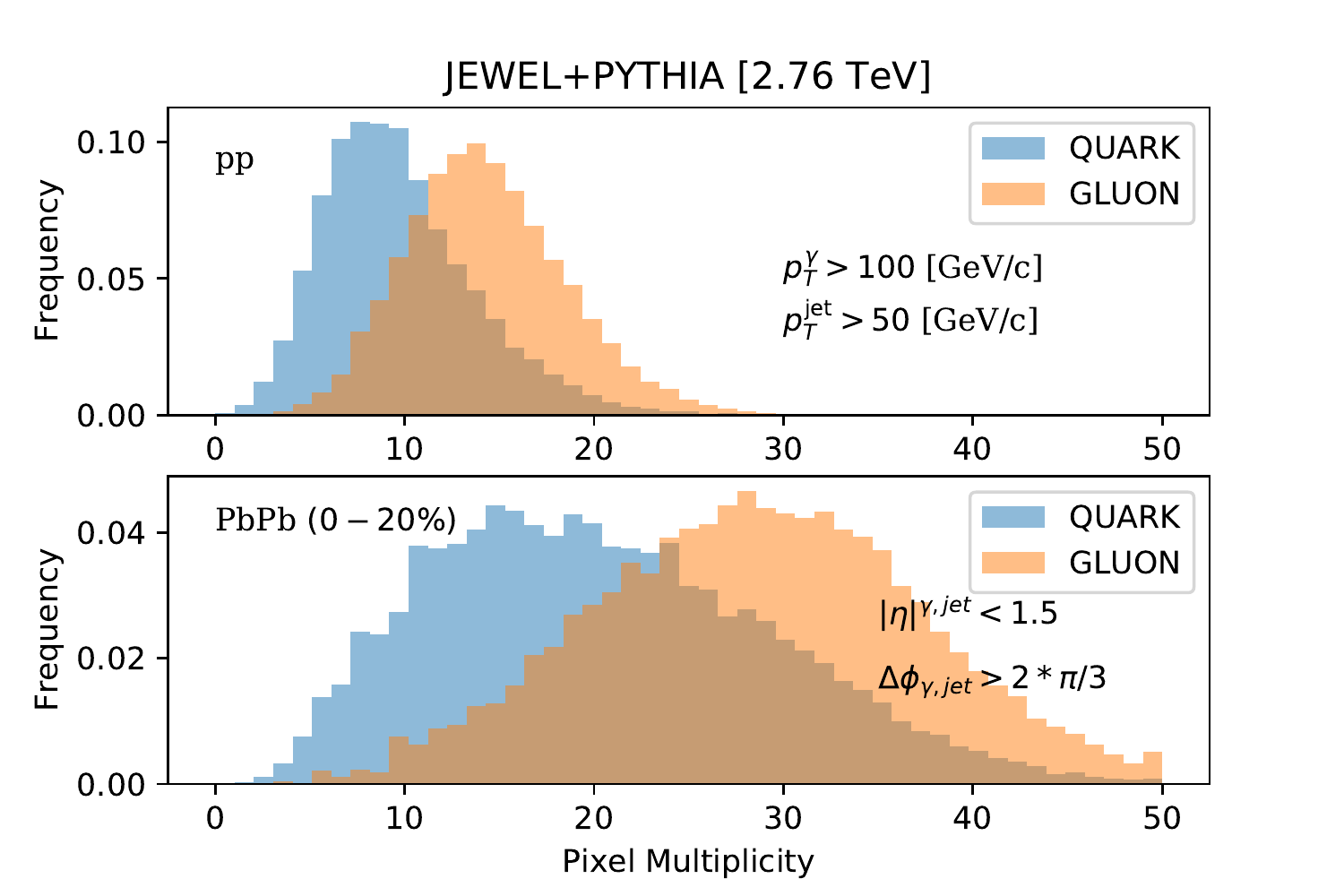}
	   \caption{Distributions of the jet mass (top left), the girth or the first radial moment (top right), the half radial moment (center left), the $p^{D}_{T}$ (center right) and the pixel multiplicity (bottom) for quark (darker blue) and gluon (lighter orange) jets. The top panels of each individual distribution show the pp ones while the bottom panels correspond to central (0-20\%) PbPb collisions generated by \jw simulations.}
	   \label{fig:jetdistributons_pp_pbpb}
	\end{figure}

Multivariate analyses have been successfully employed in object classification and selections in high energy physics \cite{Bhat:2010zz, 1742-6596-608-1-012058} and thus are an effective, physics-motivated baseline in our study. Given a jet reconstructed using the anti-$k_{t}$ algorithm, we choose the following set of five jet substructure observables \cite{Gallicchio:2012ez} to highlight the differences between quark jets and gluon jets,
	\begin{itemize}
		\item Jet mass: $m = \sqrt{(\sum_{i\in {\rm jet}} E_i)^{2} - (\sum_{i\in {\rm jet}} \vec{p_i})^2}$.
	        \item First radial moment (girth): $\sum_{i \in {\rm jet}} p_{T_i} \Delta R_{{\rm jet}, i}/p^{\rm jet}_{T}$, where $\Delta R_{{\rm jet}, i}=\sqrt{\Delta \eta^2_{{\rm jet}, i}+\phi^2_{{\rm jet}, i}}$.
		\item 0.5 radial moment:  $\sum_{i \in {\rm jet}} p_{T_i} (\Delta R_{{\rm jet}, i})^{0.5}/p^{\rm jet}_{T}$.
        		\item $p_{T}^{D}$: $p^{D}_{T} = \sqrt{\sum_{i \in {\rm jet}} {p^2_{T_i}}}/p^{\rm jet}_{T}$.
       		\item Pixel multiplicity: the number of $(\eta,\phi)$ pixels with nonzero energy deposit.
	\end{itemize}

Figure~\ref{fig:jetdistributons_pp_pbpb} shows the distributions of the aforementioned jet observables in pp and PbPb collisions for quark (blue shaded histograms) and gluon jets (orange shaded histograms). Both pp and PbPb jets include the grid discretization. Furthermore, jets in PbPb collisions are subtracted using the GridSub background subtraction method. Note that the jet mass and radial moments are infrared and collinear (IRC) safe observables with different weights on the particle angular distributions. In vacuum, they reflect the fact that gluon jets have larger jet mass and larger radial moments than quark jets due to the larger color charge and Casimir color factor $C_A>C_F$. In the \jw PbPb simulations, both quark and gluon jets have distributions that are modified towards larger values. On the other hand, we see a significant increase of the pixel multiplicity and decrease of the values of $p_T^D$ for quark and gluon jets, respectively, with the caveat that pixel multiplicity is unreliable in a heavy ion environment and the $p_T^D$ is an IRC unsafe observable.

It is important to note that these observables are sensitive to the jet clustering algorithm, the minimum \pt ~cutoff of jet constituents and the background subtraction procedures employed. In order to facilitate a direct comparison between experimental data and Monte Carlo simulations, one has to either unfold the detector resolution from these observables or perform Monte Carlo simulations with the appropriate detector response. The unfolding procedure for jet observables increases in dimensionality and complexity as the correlations among observables become strong. For example, the jet mass will have to be unfolded with a 4-dimensional response matrix consisting of the generated and reconstructed jet \pt ~and mass.

We combine the five jet substructure observables using a multivariate model implemented in Keras \cite{keras}, with two hidden dense layers of 10 nodes each. We use the rectified linear unit (ReLU) \cite{nair2010rectified} and sigmoid activation functions for the dense layers and the output layer, respectively. The model training utilizes the Adam optimizer \cite{adam} and the binary cross entropy loss function. We perform cross validation using Monte Carlo samples which are split into two random halves for training and testing.

\subsection{Jet Image}
\label{sec:image}

\begin{figure}[t]
\centering
\includegraphics[width=0.45\textwidth]{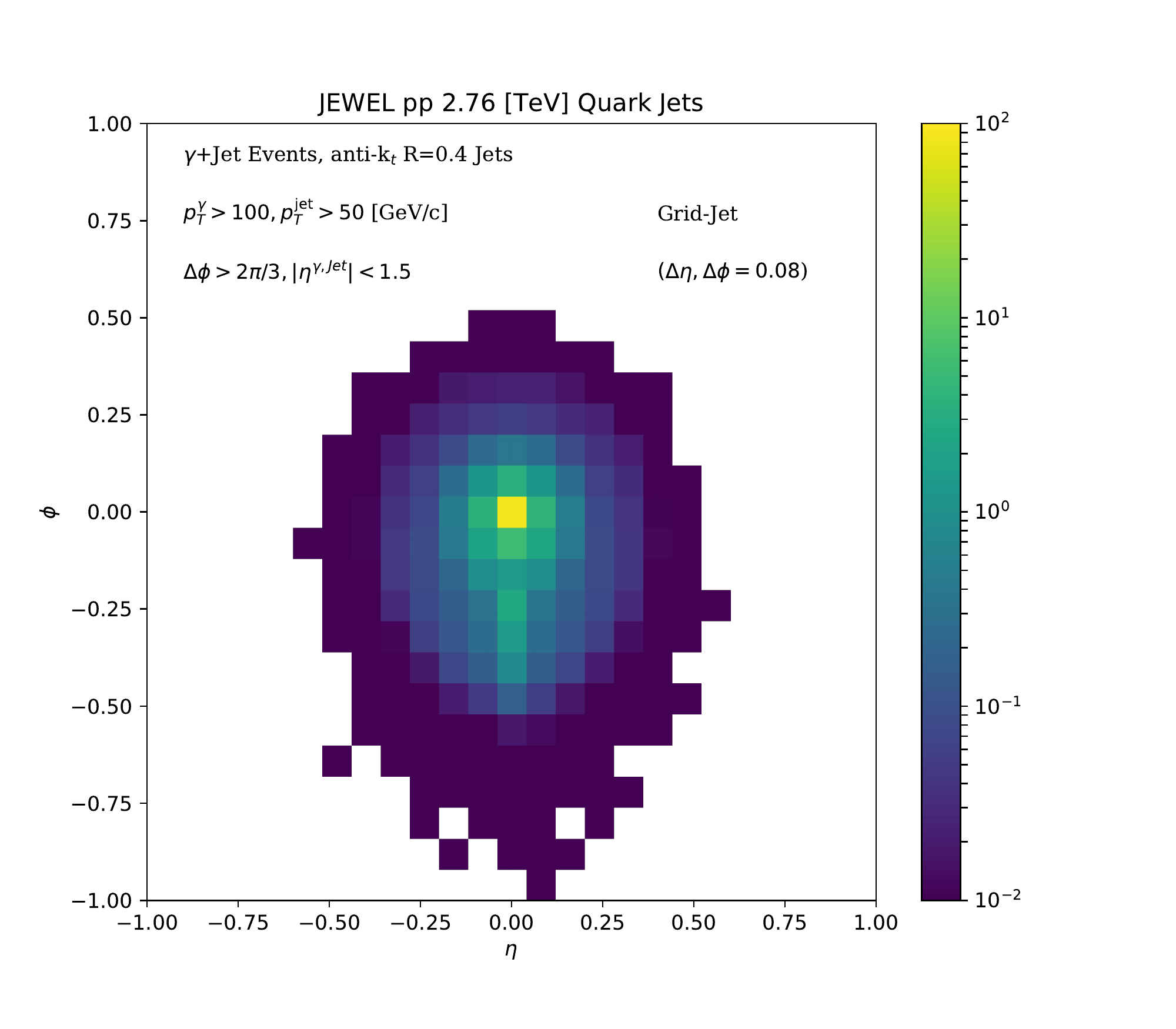}
\includegraphics[width=0.45\textwidth]{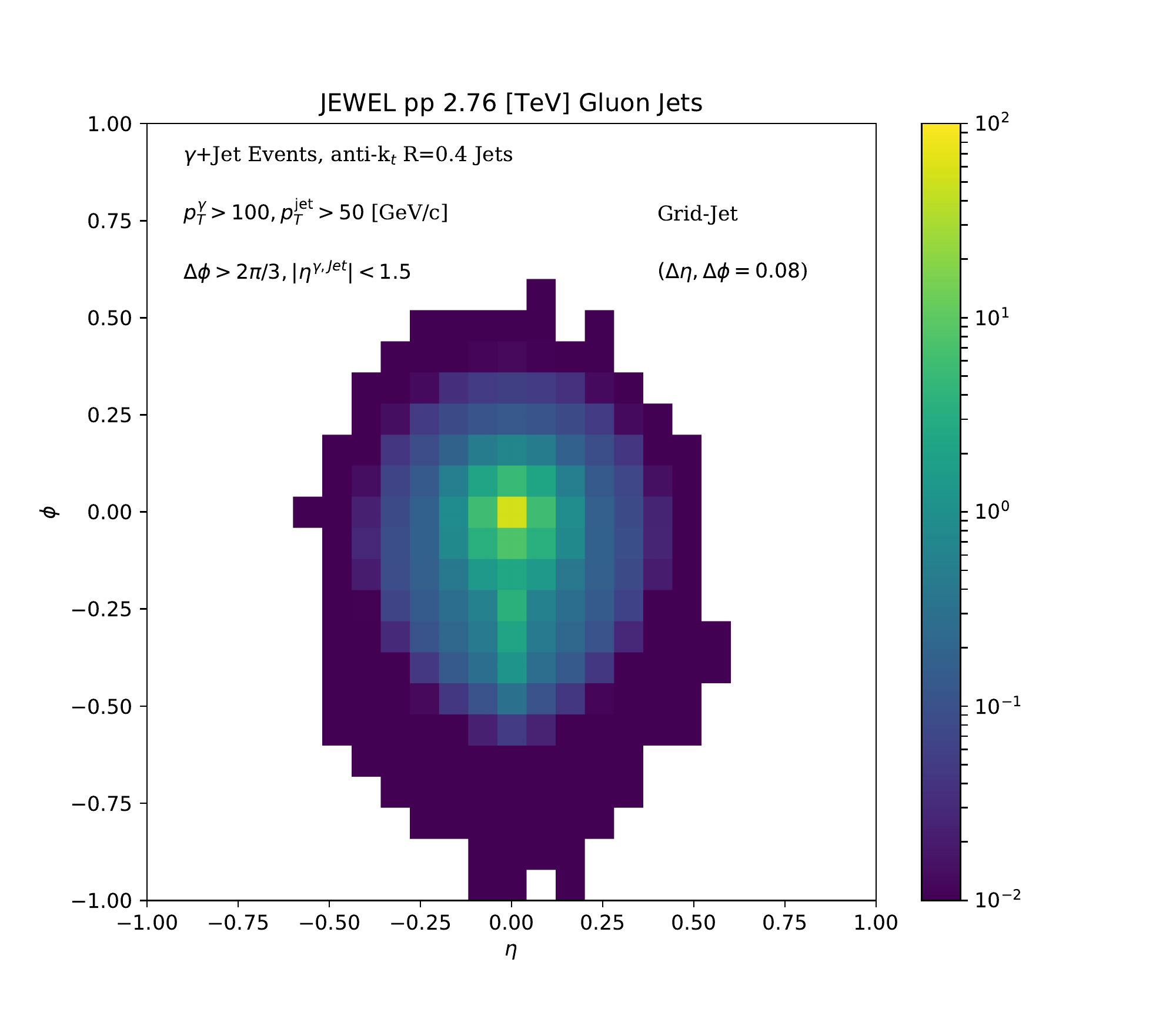}
\includegraphics[width=0.45\textwidth]{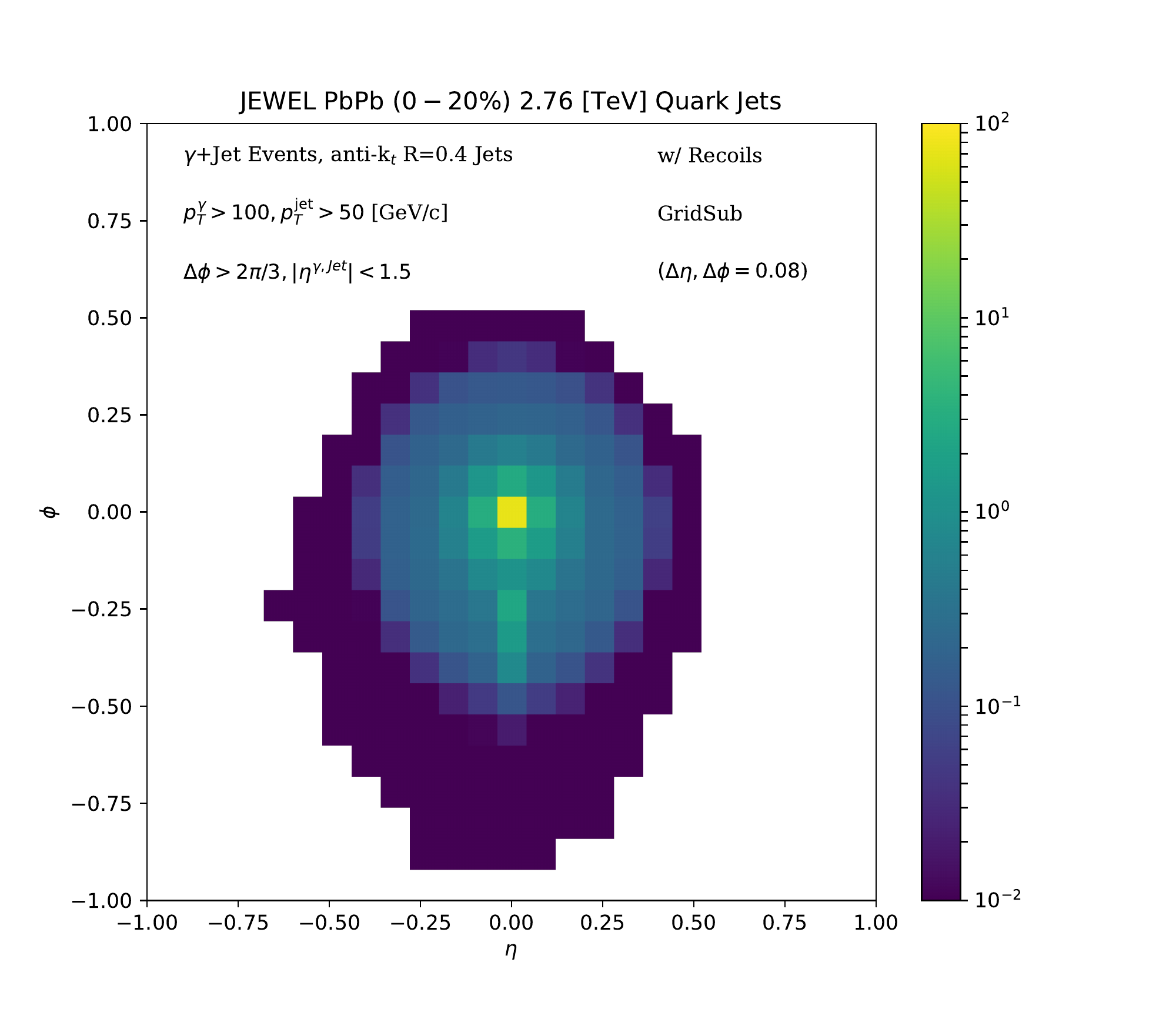}
\includegraphics[width=0.45\textwidth]{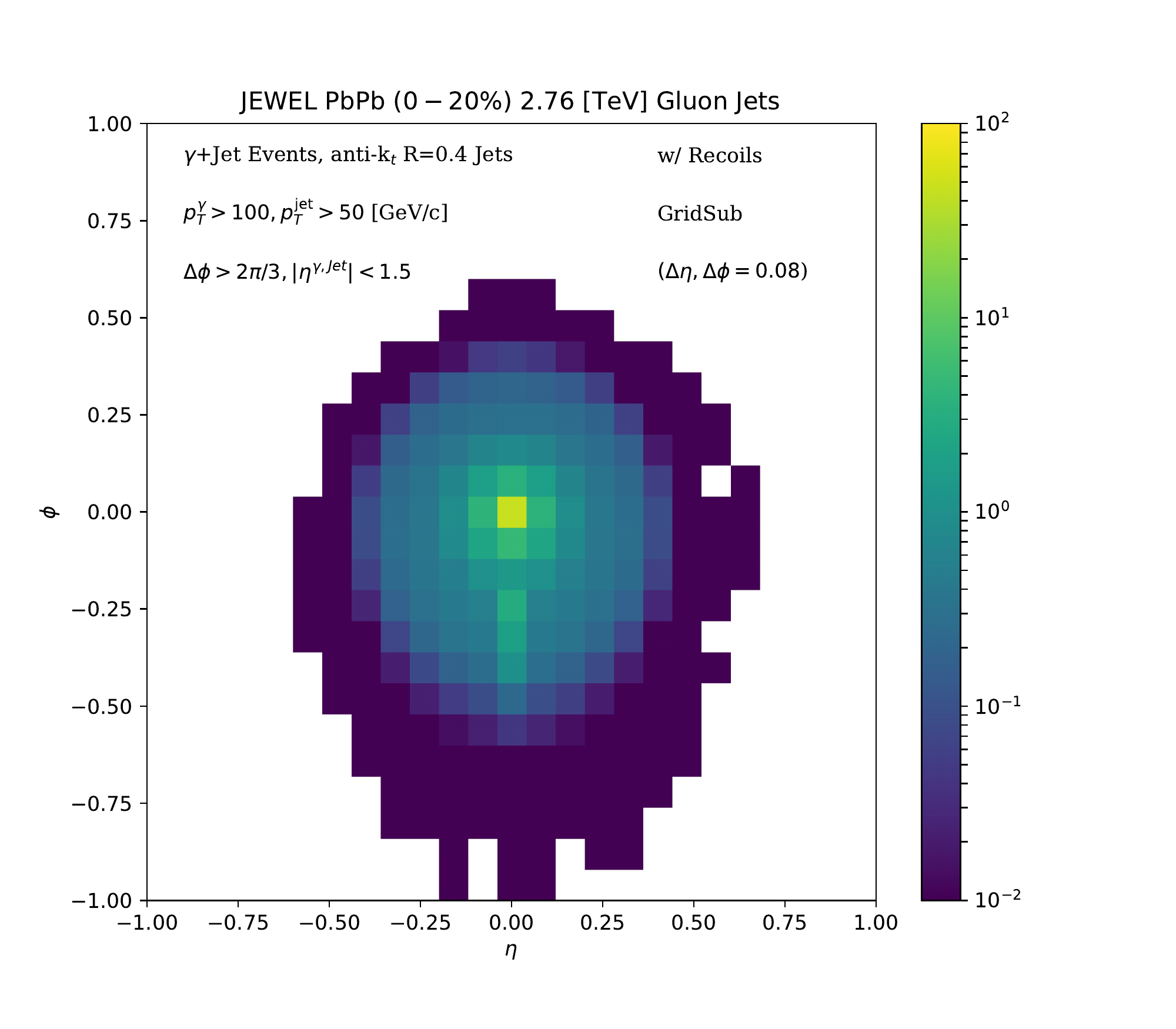}
\caption{Average, pre-processed quark (left) and gluon (right) jet images in the $\eta-\phi$ plane for pp (top) and central PbPb (bottom) collisions generated from \textsc{Jewel}. The color scale represents the average transverse momenta deposited in each pixel in units of GeV/c. }
\label{fig:qgjetimages}
\end{figure}

The multivariate analysis discussed previously is a constructive approach to collect and examine the usefulness of jet features that may help distinguish quark jets from gluon jets, or jets in different collision systems. An alternative approach is to use image-recognition techniques which optimize comprehensive multivariate models and identify all possible useful features. The radiation pattern of a jet can be quantified as an image in the $\eta-\phi$ plane with $p_T$ deposited in detector cells, much like a digital camera. The jet image thus provides a fixed-dimensional representation of jet energy distribution. Each jet image is pre-processed \cite{deOliveira:2015xxd} by translating in the $\eta-\phi$ plane so that the most active pixel is centered at the origin. Also, the image is rotated so that the principal component of the energy density lies along the $\eta=0, \phi<0$ direction.

Figure~\ref{fig:qgjetimages} shows the average quark (left) and gluon (right) jet images in pp (top) and central PbPb (bottom) collisions. The color of each image pixel represents the transverse momentum deposited in the pixel. We see a broader energy distribution around central pixels for gluon jets as compared to quark jets in pp collisions, with both flavors exhibiting significantly broader distributions for jets in PbPb collisions. Note that in experiment, it may be very challenging to directly analyze jet images due to detector and measurement considerations. Here we mainly use the jet image approach as a useful and state-of-the-art comparison of the classification performance.
	
A deep convolutional neural network is implemented in Keras with the TensorFlow \cite{DBLP:journals/corr/AbadiBCCDDDGIIK16} backend. The neural network model consists of three convolution layers of sequentially reducing sizes, each with 8 filters followed by a max-pooling layer. The convolution layers use a hyperbolic tangent activation function, and the output layer uses a sigmoid activation function. The output layer is preceded by three layers, a dropout layer with a rejection score of 0.20, a dense layer with 20 nodes and an additional dropout layer with a reduced rejection score of 0.10. The additional dropout layer helps filtering less important features from previous layers and thus increases the speed and efficiency of the model training. The model is trained using the Adam optimizer with binary cross entropy loss function.

\subsection{Telescoping Deconstruction}
\label{sec:tjet}

\begin{figure}[t]
\centering
\includegraphics[width=.49\columnwidth]{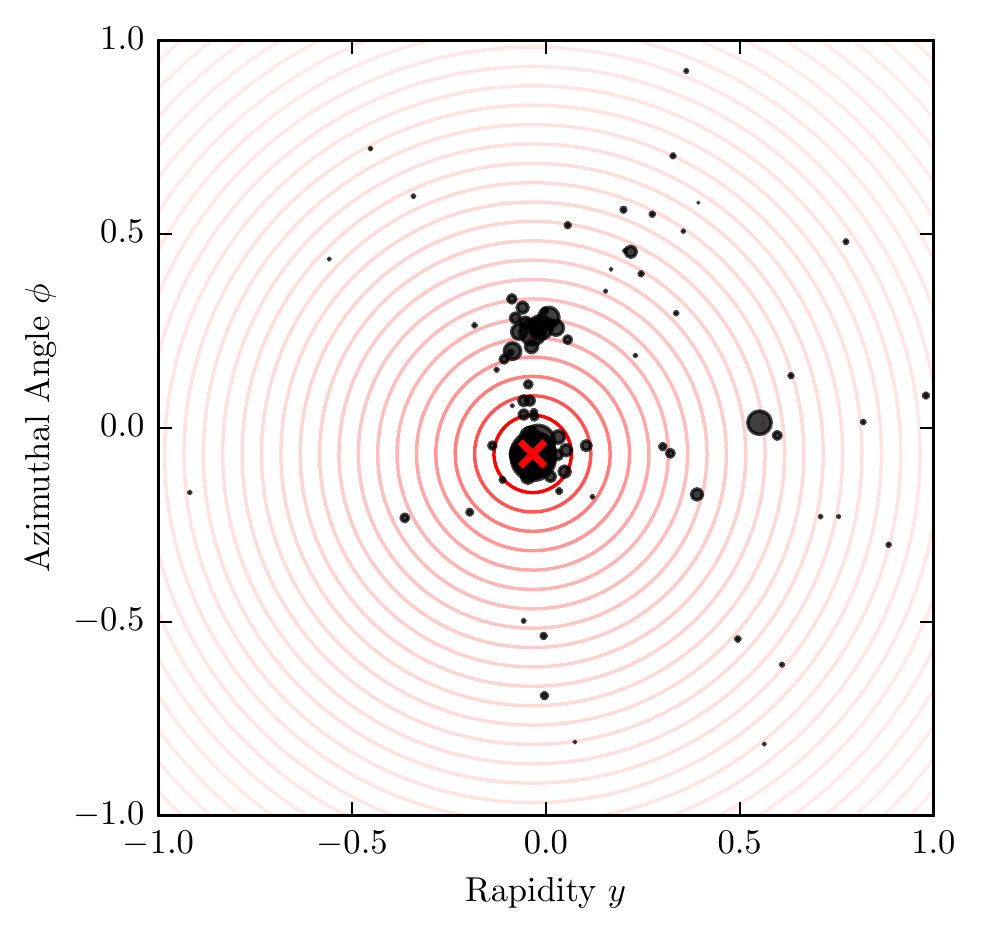}
\includegraphics[width=.49\columnwidth]{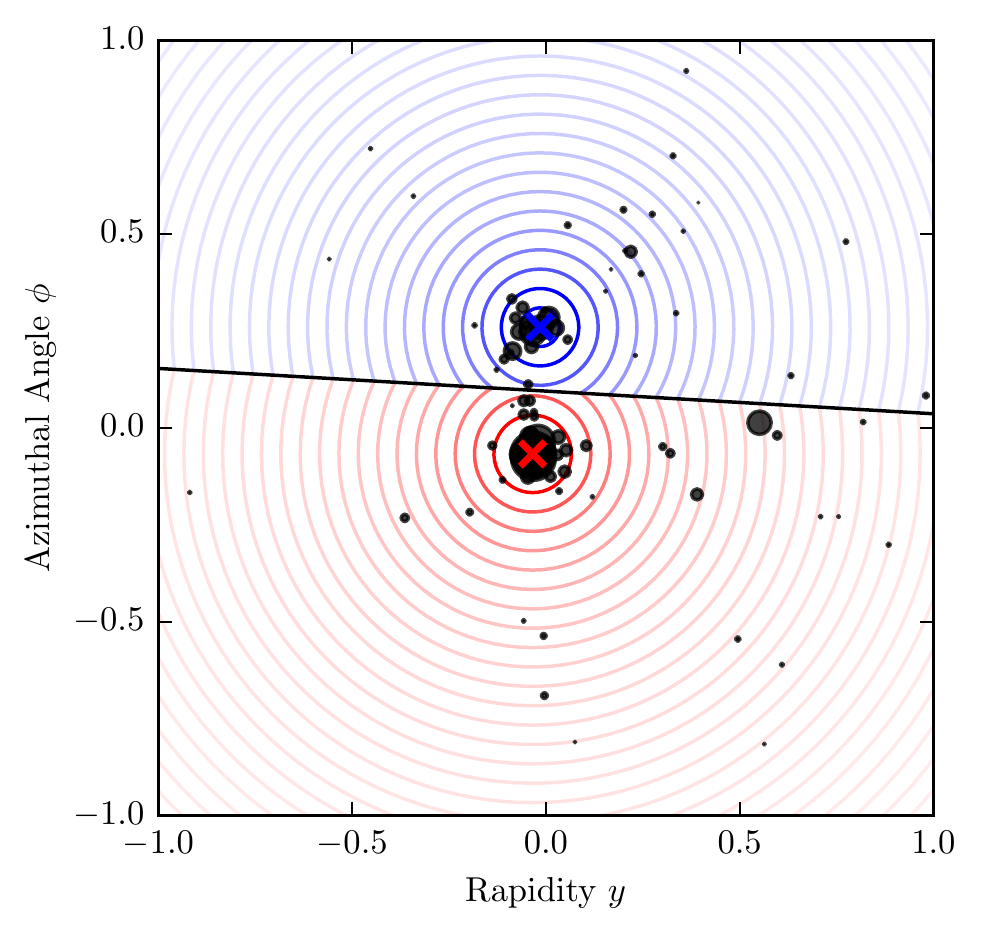}
\includegraphics[width=.49\columnwidth]{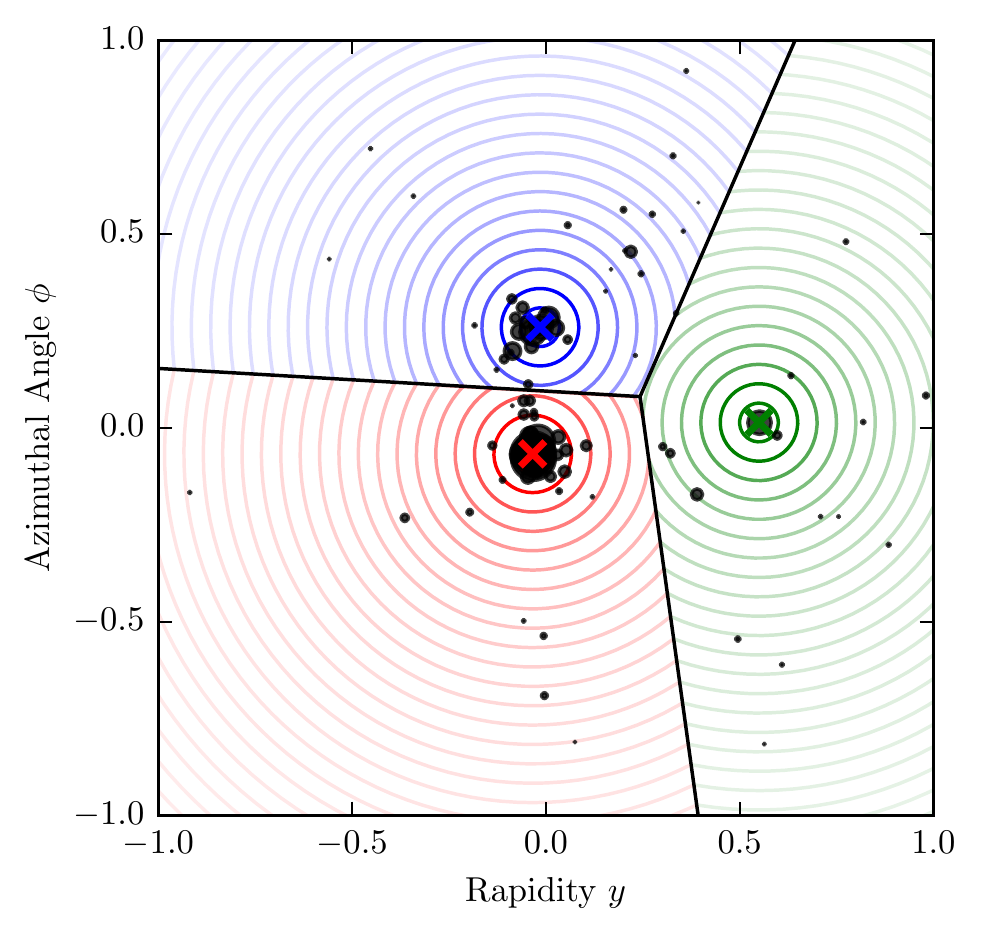}
\caption{Illustration of telescoping deconstruction at T1 (top left panel), T2 (top right panel) and T3 (bottom panel) orders for a random pp jet. The crosses are the Winner-Take-All $k_t$ axes. The straight lines are the exclusive subjet boundaries. Particles are sized according to their transverse momenta. See Appendix for more details about the procedures.}
\label{fig:tjetQCD}
\end{figure}

The multivariate analysis and jet image recognition discussed previously are two characteristically different methods. The former uses physics-motivated observables and is a ``bottom-up" approach to collect useful jet features. However, the set of observables identified may be highly correlated, and it is not clear how to systematically improve the method. On the other hand, jet images represent low-level, comprehensive jet information. The training of a deep neural network is a ``top-down" approach to identify all useful features through model optimization. However, extracting physical messages from the trained neural network parameters can be a challenging task.

We develop and apply the telescoping deconstruction (TD) framework in our study of quark and gluon jet substructures in pp and PbPb collisions. TD aims to embrace both advantages of multivariate analysis and jet image recognition. It decomposes jets using physics-motivated, comprehensive sets of observables which form an organized fragmentation basis of jet substructures. It is a complete and systematic subjet expansion~\footnote{Other useful basis of jet substructure observables have also been explored recently~\cite{Datta:2017rhs, Komiske:2017aww}.} consisting simply of subjet kinematic variables, i.e. the subjet $p_T$ and mass. The expansion is ordered by $N$, the number of subjets exclusively reconstructed, and the individual TD observables are physically motivated to encode the hard splitting of jets and relevant non-perturbative physics within each subjet variable. The telescoping procedure probes radiation around dominant energy flow in a jet with multiple angular resolutions \cite{Chien:2013kca,Adams:2015hiv,Chien:2014hla,Chien:2017xrb}. It efficiently quantifies the radiation distribution by first capturing the dominant energy in the subjet reconstruction followed by reaching out to include wide-angle, soft radiation. Figure~\ref{fig:tjetQCD}\footnote{We thank Patrick Komiske and Eric Metodiev for making the beautiful, ``{\sl Ripples in Jets}" illustration in Figure~\ref{fig:tjetQCD} and the one in the Appendix, celebrating the Nobel prize awarded to the detection of gravitational waves, i.e., {\sl Ripples in Spacetime}.} illustrates the telescoping deconstruction of a random pp jet at T1, T2 and T3 orders (clockwise from top left). Details about the telescoping deconstruction procedures are provided in the Appendix, including the applications to boosted $W$ and top tagging in high energy physics.

For $R=0.4$ jets, we telescope around the subjets axes using radii from 0.08 to 0.4 with $N_{\rm steps} = 5$ steps of 0.08. We consider $N=\{1,2,3\}$ in this study and find the performance quickly saturates at $N=3$. Each telescoping deconstruction order has $N_{\rm steps} \times 2$ variables of subjet transverse momenta and subjet masses, in addition to the subjet axis information $\eta, \phi$. 
These telescoping variables are cumulatively combined and input to a multi-layer perceptron, much like the one we used in the previous multivariate analysis, with two hidden dense layers and a final output layer with the same activation and optimization functions.

Beside the overall classification performance using telescoping deconstruction we present in the next section, subjet variables within this framework can reveal many aspects of jet dynamics. In particular, there is rich information contained in subjet topology and subjet substructure. Recently, the CMS~\cite{Sirunyan:2017bsd} and STAR~\cite{Kauder:2017mhg} collaborations measured the groomed momentum sharing observable $z_g$. Jets are reclustered using the Cambridge/Aachen (C/A) algorithm \cite{Dokshitzer:1997in,Wobisch:1998wt} and wide angle, soft radiation are sequentially removed until the following condition is satisfied by a branching,
\begin{equation}
    z_{\rm cut} < \frac{\min(p_{T_1},p_{T_2})}{p_{T_1}+p_{T_2}} \equiv z_g\;.
\label{SD}
\end{equation}
Here $p_{T_1}$ and $p_{T_2}$ are the transverse momenta of the two branches, and $z_{\rm cut}$ is the soft-drop parameter which is set to $0.1$ in our analysis. Note that the C/A clustering tree is angular-ordered and therefore the above procedure identifies the most wide angle, soft subjet that carries a significant fraction of the jet transverse momentum larger than $z_{\rm cut}$. The sizes of the subjets are dynamically determined and related to the angle $r_g$ between the two soft-drop branches, defined as the groomed jet radius. The observables $z_g$ and $r_g$ thus encode the kinematic information of the two subjets which is in the category of subjet topology. In vacuum, the observable $z_g$ is closely related to the Altarelli-Parisi splitting functions~\cite{Altarelli:1977zs}. The physical meaning of $z_g$ in heavy ion collisions have also been actively investigated \cite{Chien:2016led,Mehtar-Tani:2016aco,Milhano:2017nzm,Chang:2017gkt,Li:2017wwc}.

\begin{figure}[t]
	   \centering
	   \includegraphics[width=0.9\textwidth]{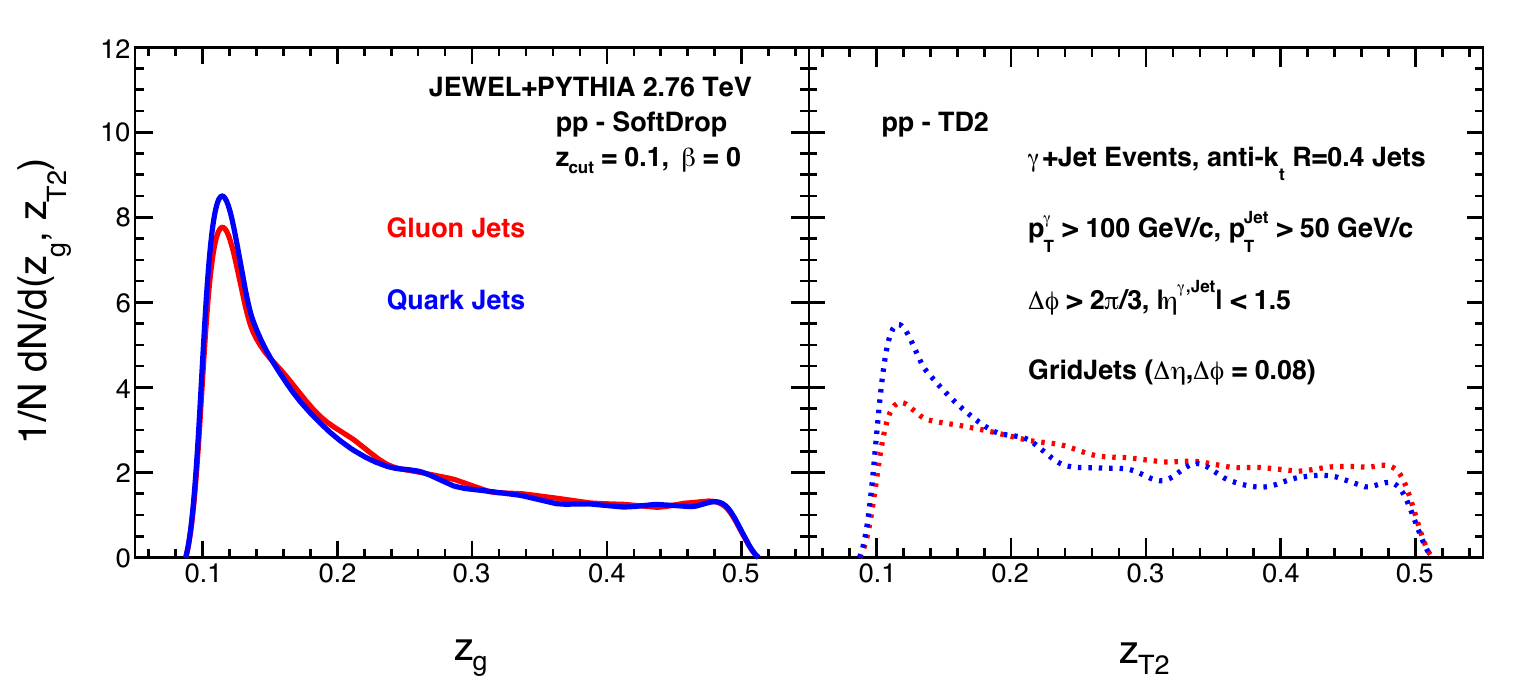}
	   \includegraphics[width=0.9\textwidth]{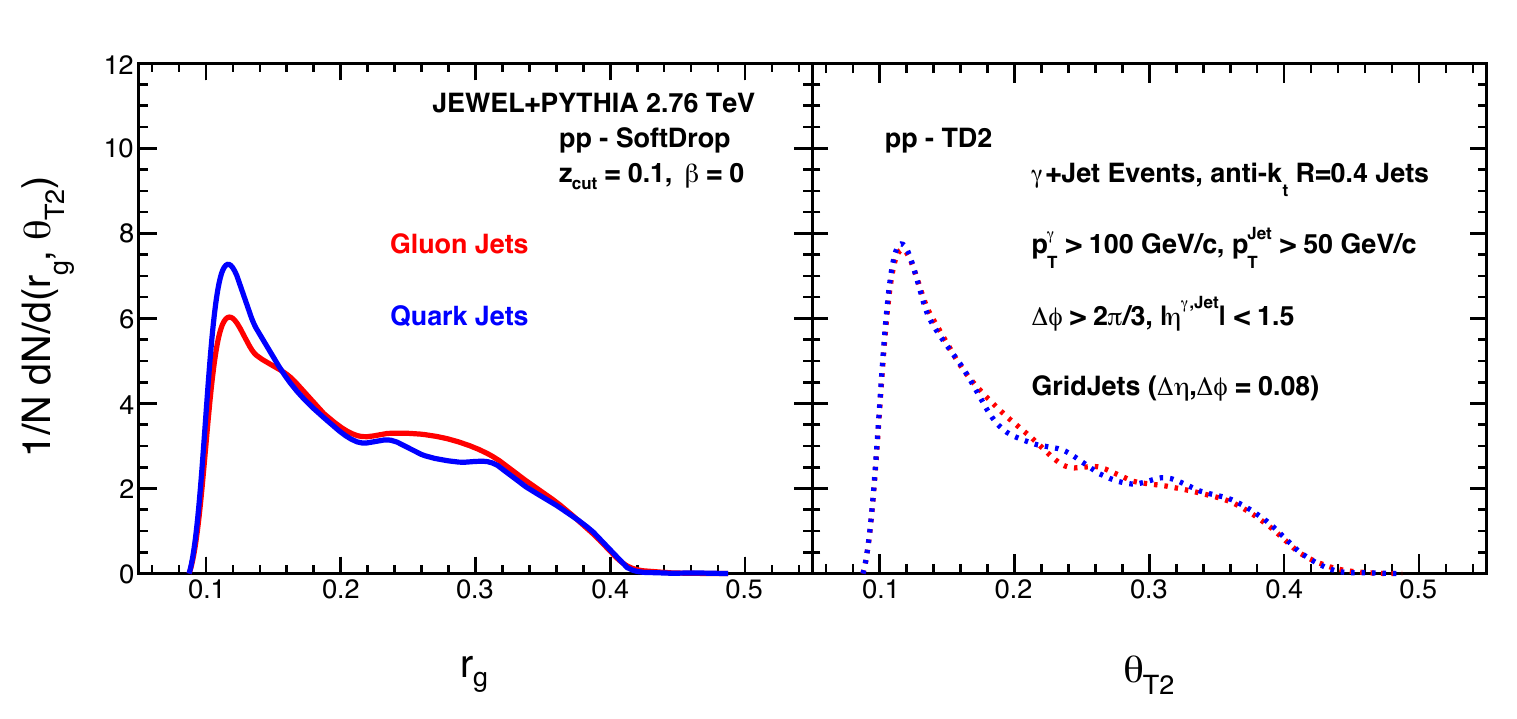}
	   \caption{Top panels: Distributions of groomed momentum sharing $z_g$ (left) and the T2 order TD subjet momentum fraction $z_{\rm T2}$ (right); Bottom panels: Distributions of groomed jet radius $r_g$ and the T2 order TD angular separation between subjets $\theta_{\rm T2}$ for quark (blue) and gluon (red) jets in pp \jw simulations.}
\label{fig:comp_z_pp_wT2}
\end{figure}

\begin{figure}[t]
	   \centering
	   \includegraphics[width=0.9\textwidth]{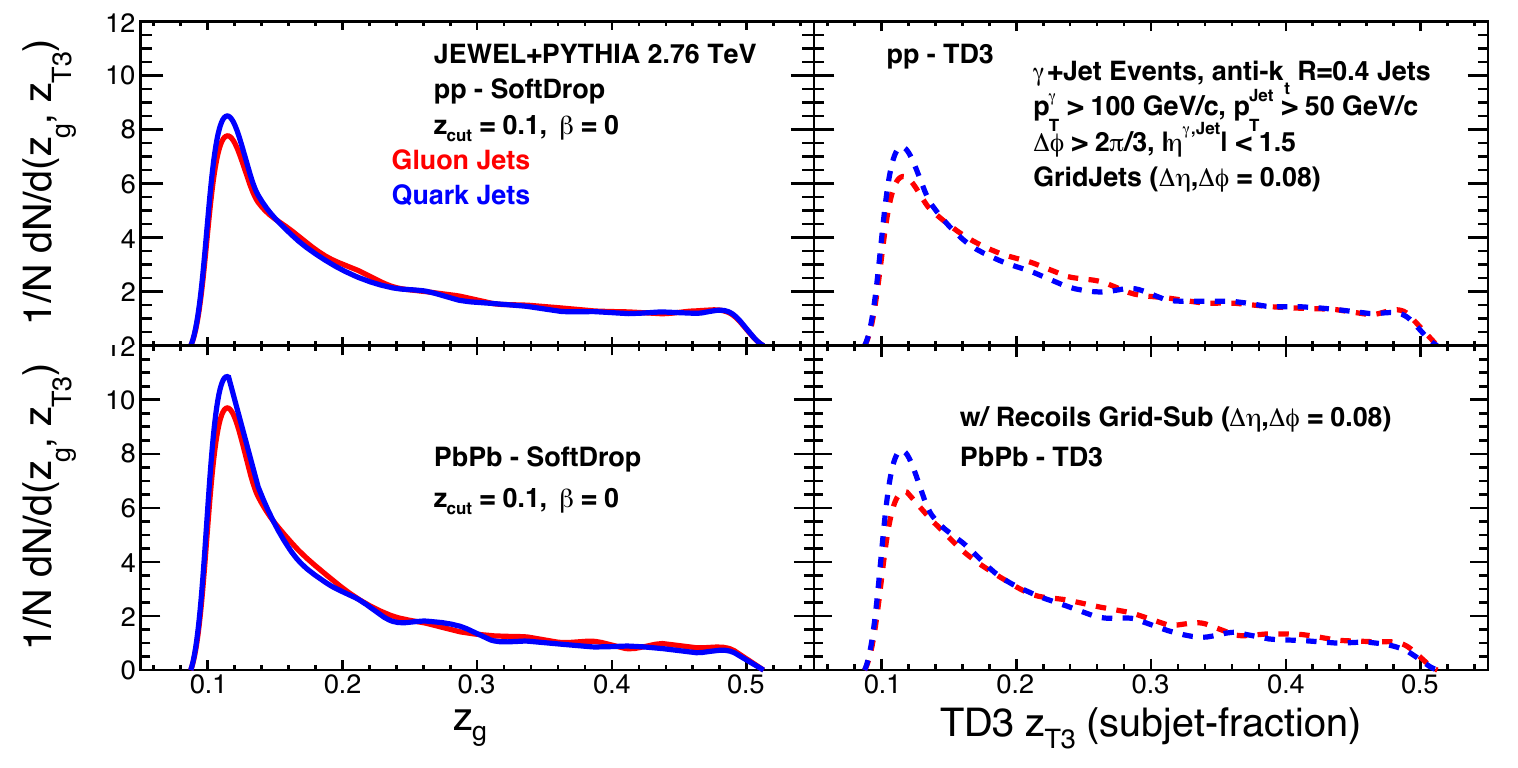}
	   \caption{Distributions of groomed momentum sharing $z_g$ (left) and TD subjet momentum fraction $z_{\rm T3}$ (right) of quark (blue) and gluon (red) jets in pp (top panels) and central PbPb (bottom panels) collisions in \textsc{Jewel}. Note that $z_{\rm cut}<z_g,z_{\rm T3}<0.5$ and $z_{\rm cut}$ is chosen to be 0.1 in this work. (color online)}
\label{fig:comp_z}
\end{figure}

We use telescoping deconstruction to analyze the two soft-drop branches that define the $z_g$ observable. As a first attempt, at the T2 order we choose the subjet radius $r$ proportional to the angle $\theta$ between the two deconstruction axes for subjet reconstruction: $r=C \theta$. The proportionality constant $C$ is set to be 0.3, to force the two subjets separated in angle. We impose a similar condition for the momentum fraction $z_{\rm T2}$ of the softer subjet to be larger than $z_{\rm cut}$,
\begin{equation}
    z_{\rm cut} < \frac{\min(p_{T_1},p_{T_2})}{p_{T_1}+p_{T_2}} \equiv z_{\rm T2}\;,
\label{T2_z}
\end{equation}
where $p_{T_1}$ and $p_{T_2}$ are the transverse momenta of the two exclusively constructed subjets at the T2 order. The resulting $z_{\rm T2}$ distribution is quite different from the $z_g$ distribution, especially for gluon jets as shown in the top panels of Figure~\ref{fig:comp_z_pp_wT2}.  Note that the axes used in the telescoping deconstruction procedure are defined with the winner-take-all scheme which favors energetic particles. We find that the axes determined at the T2 order may both align with the hard soft-drop branch and are not able to efficiently tag the soft branch when it consists mainly of soft particles.

The situation is significantly improved at the T3 order where three axes $\{\hat n_1,\hat n_2,\hat n_3\}$ are used to capture the soft branch in soft-drop. Similarly, the subjet radii $r_i$ for the $i$-th subjet are chosen to be proportional to the angles among the three axes: $r_i = C \max_{j} \{\theta_{ij}\}$ where $\theta_{ij}=\cos^{-1}(\hat n_i\cdot\hat n_j)$. We combine the two closest subjets and construct the momentum fraction of the softer of the resulting two subjets with a similar kinematic cut,
\begin{equation}
    z_{\rm cut} < \frac{\min(p_{T_1},p_{T_2})}{p_{T_1}+p_{T_2}} \equiv z_{\rm T3}\;,
\label{T3_z}
\end{equation}
as well as the angle between the two subjets $\theta_{\rm T3} > \Delta$ where $\Delta$ is chosen to be 0.1 \cite{Sirunyan:2017bsd}.
This identifies the kinematics of the branching at the root of the C/A clustering of the three telescoping subjets. The distributions of the momentum fraction $z_{\rm T3}$ for quark (blue) and gluon (red) jets are very similar as shown in Figure~\ref{fig:comp_z}.
The left panels represent the soft-drop $z_{g}$ distributions in pp and PbPb collisions, and we see an enhancement of low-$z_g$ jets in the \jw PbPb collisions \cite{KunnawalkamElayavalli:2017hxo,Milhano:2017nzm}. On the other hand, the momentum fraction $z_{\rm T3}$ (right panels) for both quark and gluon jets has a similar enhancement at low $z_{\rm T3}$ values, but its distribution is less modified by quenching.

\begin{figure}[t]
	   \centering
	   \includegraphics[width=0.9\textwidth]{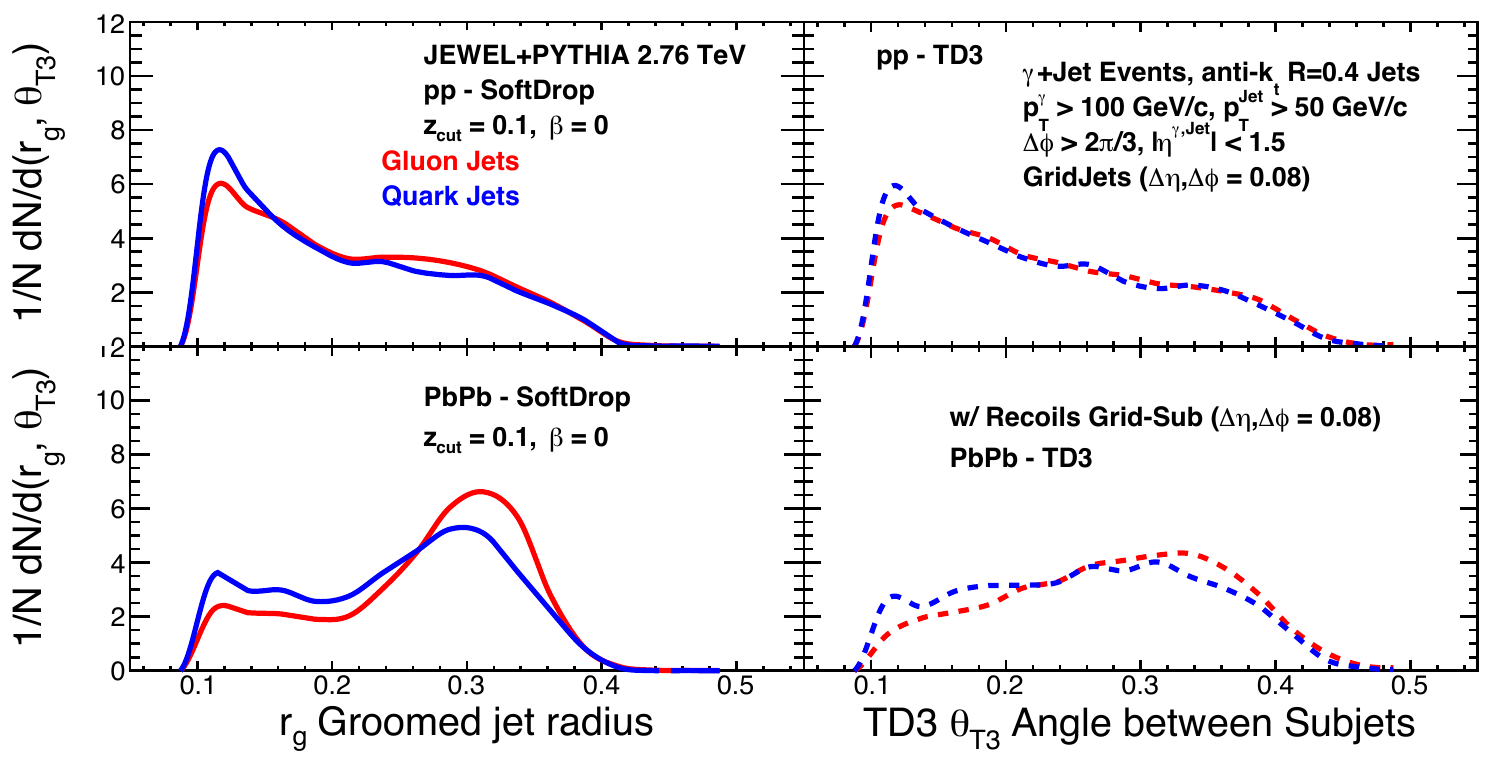}
	   \caption{Distributions of groomed jet radius $r_g$ (left) and TD angular separation between subjets $\theta_{\rm T3}$ (right) of quark (blue) and gluon (red) jets in pp (top panels) and central PbPb (bottom panels) collisions in \textsc{Jewel}. The quenched $r_{g}$ distribution is significantly modified with enhancement at the edge of jets.}
\label{fig:comp_rg}
\end{figure}

In addition to the subjet momentum fraction which provides information about the longitudinal momentum distribution, the angular separation between the two subjets encodes information about the momentum distribution transverse to the jet direction. Figure~\ref{fig:comp_rg} shows the distributions of the groomed jet radius $r_g$ (left panels) and the angle $\theta_{\rm T3}$ (right panels) for quark (blue) and gluon (red) jets. 
Both the observables have similar distributions for quark and gluon jets in vacuum as shown in the top panels of Figure~\ref{fig:comp_rg}. However, we see a significant enhancement of the $r_g$ distribution at large angle ($r_g\approx0.3$) near the edge of the quenched jets, while the modification of $\theta_{\rm T3}$ has a similar enhancement but with smaller magnitude (bottom panels). Note the difference in the subjet axis definitions which affects the angles among them: soft-drop uses C/A axes whereas TD uses WTA axes. Also, the subjet radii in TD provide an extra handle on the area of coverage by subjets therefore soft radiation can be probed with various angular resolutions. While these observables have not yet been experimentally measured at the RHIC or the LHC, the observed phenomenon of jet broadening in the \jw simulations is consistent with the jet shape measurement~\cite{Chatrchyan:2013kwa}.

In vacuum, we see that all the $\{z_g,r_g\}$ and $\{z_{\rm T3},\theta_{\rm T3}\}$ distributions are relatively insensitive to the partonic origin of jets because the color factors cancel in the normalization of the distributions. They encode information about the leading-order subjet topology which is determined by the QCD splitting kinematics.

\begin{figure}[t]
	   \centering
	   \includegraphics[width=0.9\textwidth]{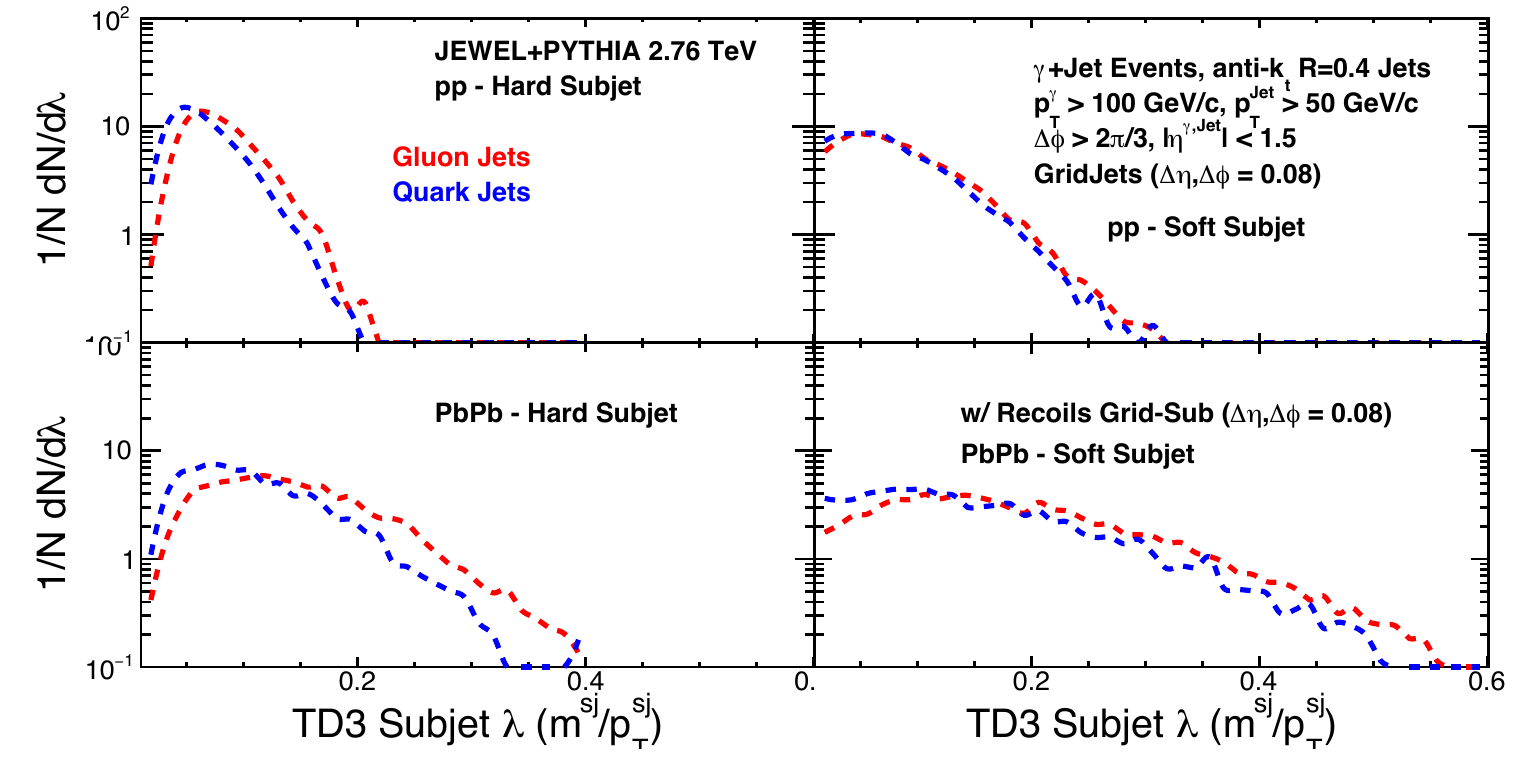}
	   \caption{Distributions of TD $\lambda$, which is the ratio between subjet mass and subjet $p_T$ for hard (left) and soft (right panels) subjets with quark (blue dashed) and gluon (red dashed) jets in pp (top) and central PbPb (bottom) collisions in \textsc{Jewel}. }
\label{fig:comp_subjet_m}
\end{figure}

To go beyond to higher-order jet features, we also study the subjet substructure, for example subjet masses. These observables disentangle the subjet kinematic information and are more directly sensitive to the soft radiation around subjets. Figure~\ref{fig:comp_subjet_m} shows the distributions of $\lambda\equiv m/p_T$, which is the ratio between subjet mass and subjet $p_T$, of hard subjets (left panels) and soft subjets (right panels) for quark (blue dashed) and gluon (red dashed) jets. Here hard or soft refers to subjet momenta with harder meaning the one with higher transverse momentum. We see hints that the $\lambda$ distributions of hard subjets exhibit flavor dependence, while those of soft subjets are similar for quark and gluon jets. This is consistent with the physics picture that for a quark jet the hard subjet is typically a quark-initiated subjet, while the soft subjet is a gluon-initiated subjet through $q\rightarrow qg$ splitting. On the other hand, both the subjets of a gluon jet are mostly gluon-initiated subjets through the $g\rightarrow gg$ splitting. We also see the effect of quenching (right panels in Figure~\ref{fig:comp_subjet_m}), where the hard and soft subjet $\lambda$ distributions get modified significantly, especially for soft subjets which tend towards larger values with an enhanced tail. This suggests that the qualitative feature of the soft event activities in the \jw simulations, persists to affect not only the subjet topology but also the subjet substructure.

\begin{figure}[t]
	   \centering
	   \includegraphics[width=0.9\textwidth]{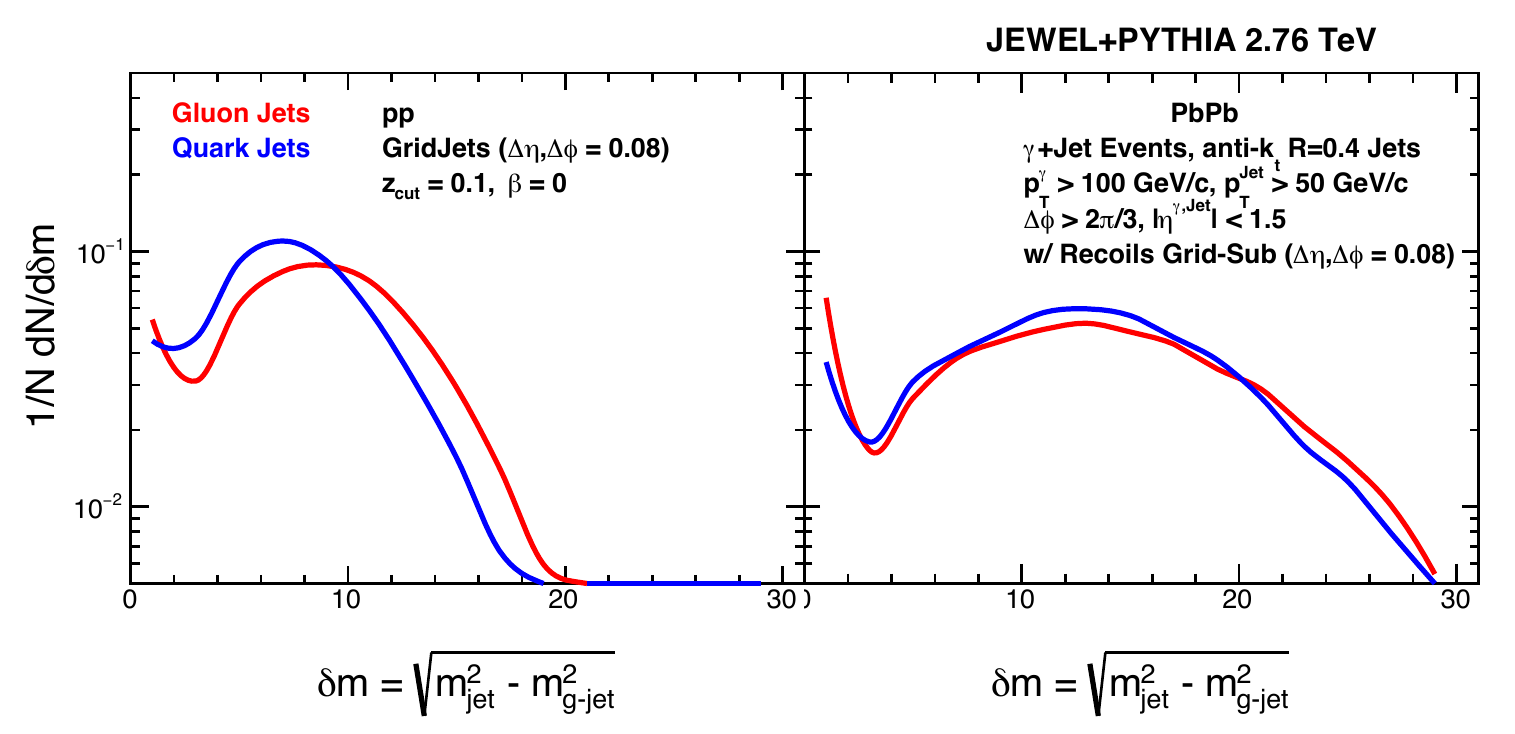}
	   \caption{Distributions of $\delta m=\sqrt{m_{\rm jet}^2-m_{\rm groomed-jet}^2}$ for quark (blue) and gluon (red) jets in pp (left) and central PbPb (right) collisions in \textsc{Jewel}.}
\label{fig:comp_delta_m2}
\end{figure}

Using the general idea of telescoping jet substructure and variability~\cite{Chien:2017xrb}, we also look at the difference between the square of jet mass for ungroomed and soft-drop groomed jets: $\delta m \equiv \sqrt{m_{\rm jet}^2-m_{\rm groomed-jet}^2}$. Since soft-drop removes soft, wide angle radiation from jets, $\delta m$ disentangles hard, collinear radiation and probes the soft radiation that is dropped. Figure~\ref{fig:comp_delta_m2} shows the $\delta m$ distributions for quark (blue) and gluon (red) jets in pp (left) and central PbPb (right) collisions simulated in \textsc{Jewel}. We see that the observable $\delta m$ in pp collisions exhibits differences between quark jets and gluon jets. On the other hand, the distributions for quark and gluon jets become very similar in PbPb collisions, albeit shifted towards higher values, and consequently become broader as compared to pp collisions. This hints that soft event activities in \textsc{Jewel} can end up smearing the differences between quark and gluon jets, which leads to the lower classification performance as we will show in the next section. Another interesting feature highlighted by the $\delta m$ distributions is the peak at zero, i.e. there is significant fraction of jets of which the masses are essentially unaffected by soft-drop. In other words, for such jets no particles are removed by soft-drop. As shown in Figure~\ref{fig:comp_delta_m2}, the distribution at $\delta m \approx 0$ is higher for gluon jets as opposed to quark jets and this particular trend persists even after quenching.

\begin{figure}[t]
	\centering
	\includegraphics[width=0.9\textwidth]{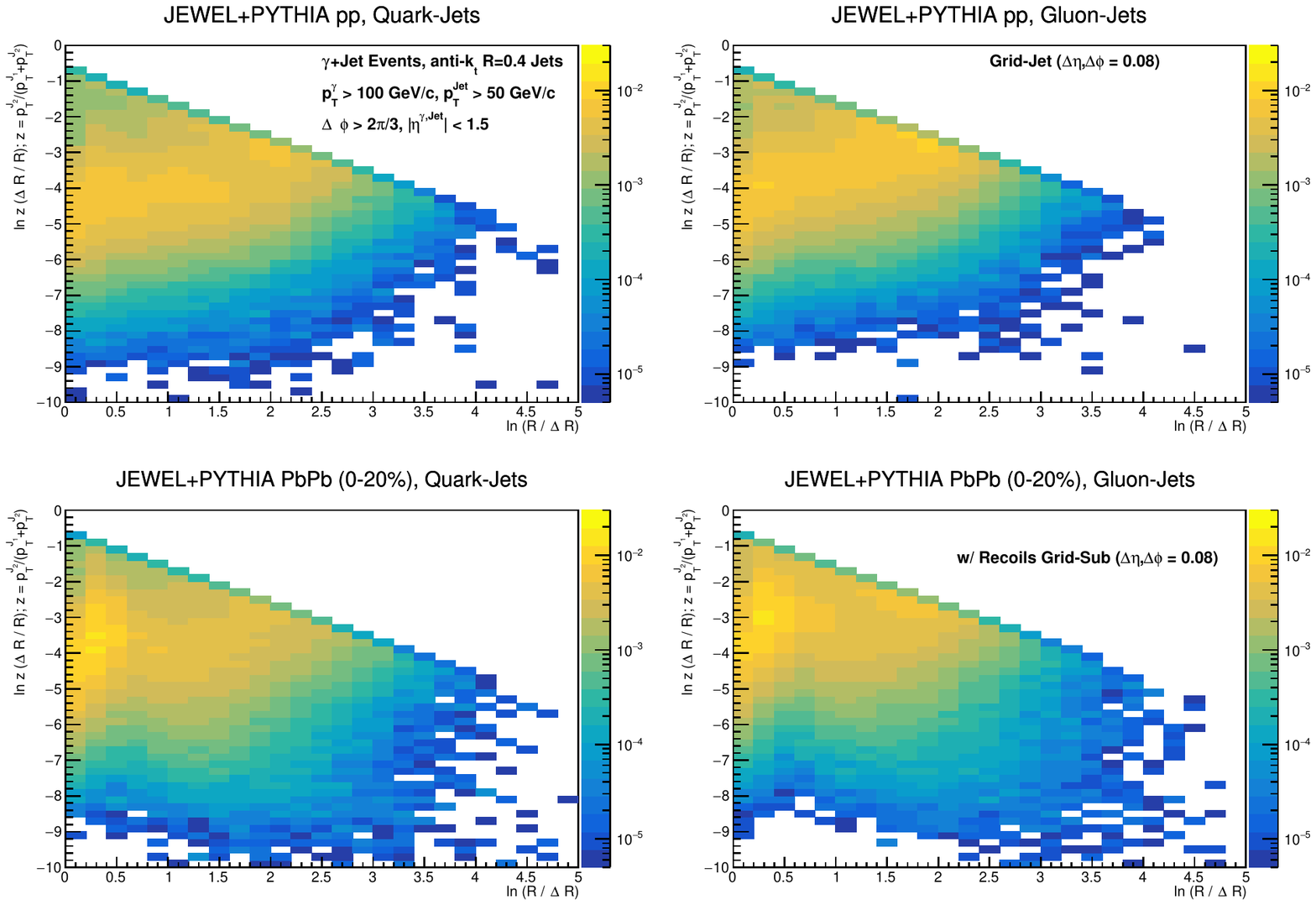}
	\caption{Lund diagrams for quark (left) and gluon jets (right) in pp (top) and central PbPb collisions in \textsc{Jewel}. See text for the definition of the axes.}
\label{fig:Lund_full}
\end{figure}

The quenching mechanism in \textsc{Jewel} probed by telescoping deconstruction variables and soft-drop can be further illuminated using the Lund diagram \cite{Andersson1989}. The Lund diagram~\cite{Andersson1989} is a diagrammatic jet representation associated with the Cambridge/Aachen clustering tree structure. It directly records the branching kinematics along the hard branches in the C/A clustering tree in terms of longitudinal momentum fraction $z$ of the soft branch and the angular separation $\Delta R$ between the two branches. The Lund diagram is then a two-dimensional histogram of variables $\ln (R/\Delta R)$ and $\ln z(\Delta R/R)$ where $R$ is the jet radius used in jet reconstruction. The variables are chosen such that the Lund diagram is uniform in the fixed strong coupling constant limit for jets in pp collisions due to the infrared structure of QCD and as such, they have been shown to be useful in distinguishing between QCD and boosted heavy particle jets \cite{Salam:2016yht}.

Given a jet reconstructed using the anti-$k_{t}$ algorithm with $R = 0.4$, the procedure of creating a Lund diagram is as follows:
\begin{itemize}
	\item Recluster the jet using the C/A algorithm
	\item Start from the root of the tree and move along the hard branches. At each branching, construct the momentum fraction of the soft branch $z=\min(p_{T_1},p_{T_2})/(p_{T_1}+p_{T_2})$ and the angular distance between the two branches $\Delta R = \sqrt{\Delta \eta^2_{1, 2}+\Delta \phi^2_{1, 2}}$
	\item Fill in the Lund diagram for each branching
	\item Continue along the subsequent hard branches until reaching only one particle
\end{itemize}

In Lund diagrams, wide angle, soft radiation populates close to the $y$-axis, while collinear, hard radiation is along the diagonal (with the negative slope). Figure~\ref{fig:Lund_full} shows the Lund diagrams for quark (left) and gluon (right) jets in pp (top) and PbPb (bottom) collisions in \textsc{Jewel}. We see a significantly enhanced region close to the $y$-axis in PbPb collisions that corresponds to increased wide angle, soft radiation in quenched jets. Comparing quark to gluon jets, we see that gluon jets (right panel) have a larger fraction of soft radiation. On the other hand, the hard, collinear radiation region of the Lund diagram is shown to reduce in its significance for quenched jets. This suggests again that jet cores are modified in \textsc{Jewel} PbPb collisions where subleading subjets tend to have wider angles and softer momenta. These are characteristic features we hope to extract, and using the Lund diagram one can directly see the emergence of such qualitative features.

\begin{figure}[t]
	\centering
	\includegraphics[width=0.9\textwidth]{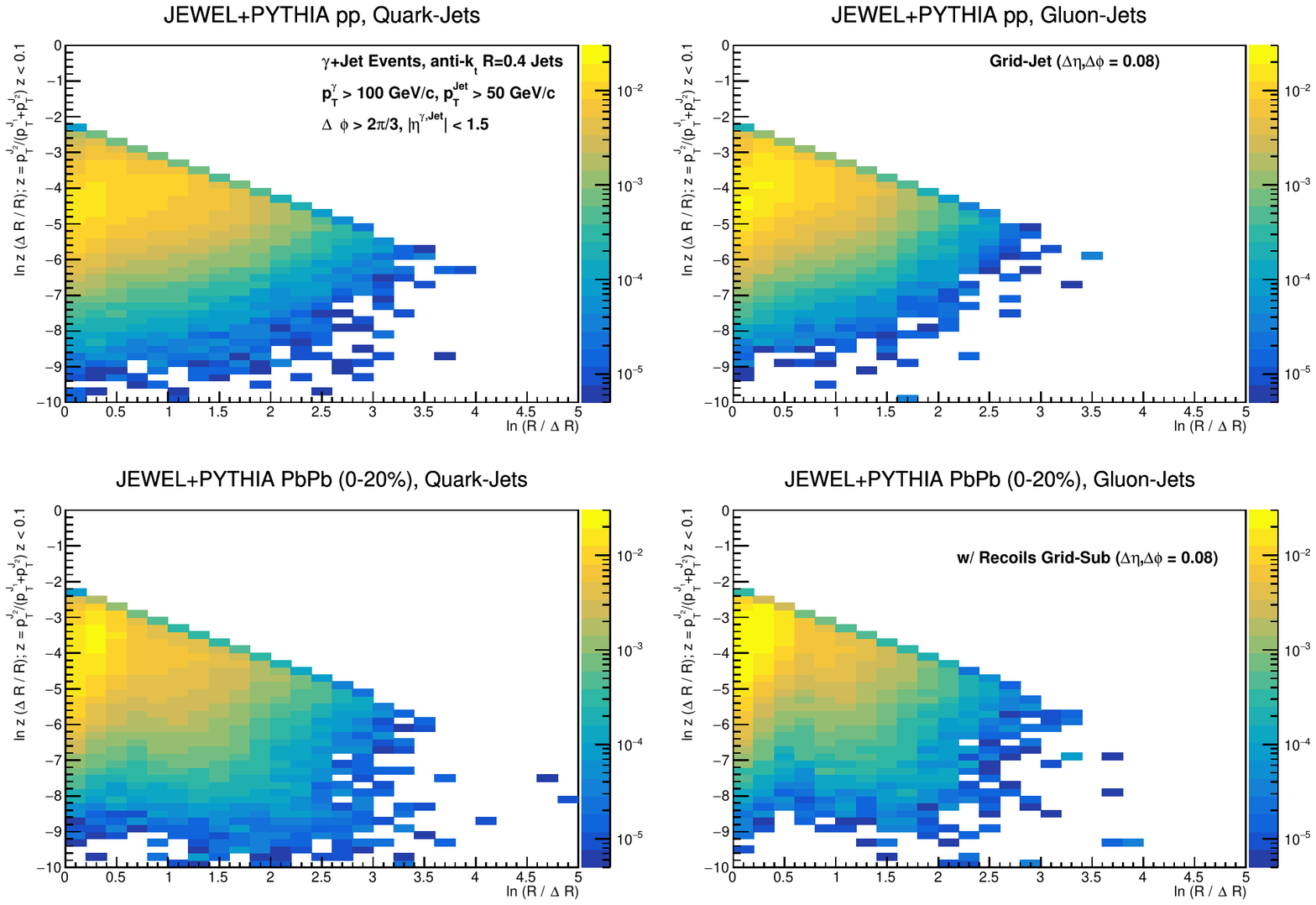}
	\caption{Lund diagrams for branches removed in the soft-drop procedure with $z<z_{\rm cut}=0.1$ for quark (left) and gluon (right) jets in pp (top) and central PbPb collisions in \textsc{Jewel}. See text for the definition of the axes.}
\label{fig:Lund_bkg}
\end{figure}

\begin{figure}[t]
	   \centering
	   \includegraphics[width=0.9\textwidth]{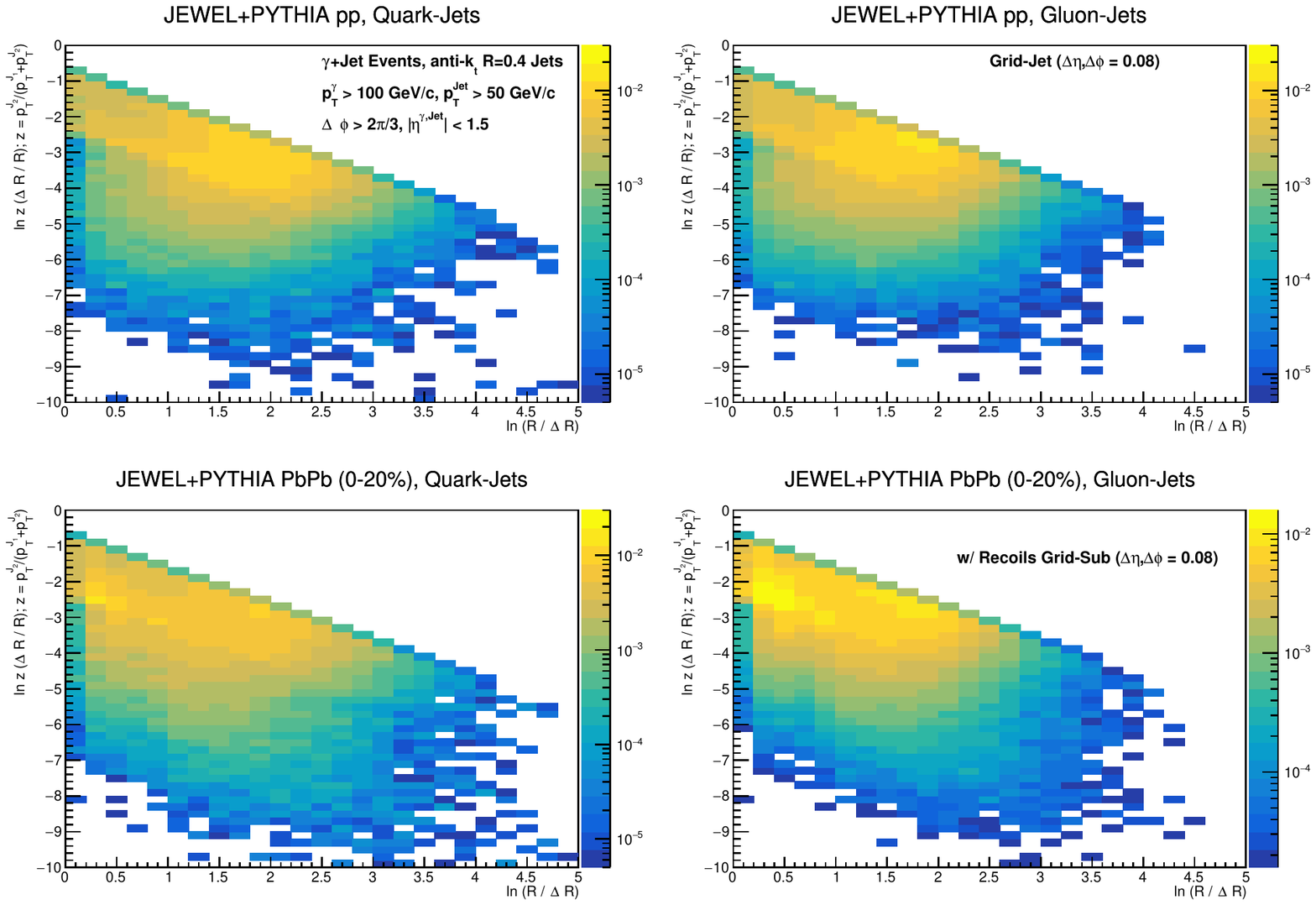}
	   \caption{Lund diagrams for branches in the hard branch after the soft-drop procedure with $z<z_{\rm cut}=0.1$ for quark (left) and gluon (right) jets in pp (top) and central PbPb collisions in \textsc{Jewel}. See text for the definition of the axes.}
\label{fig:Lund_hard}
\end{figure}

We can also study the effect of soft-drop using the Lund diagrams. Figure~\ref{fig:Lund_bkg} shows the Lund diagrams for quark (left) and gluon (right) jets in pp and PbPb collisions, with only the branchings right up to the one defining the $z_g$ and $r_g$ observables. That is, Figure~\ref{fig:Lund_bkg} corresponds to radiation that is dropped in the soft-drop procedure. Note the strict cutoff of $z < z_{\rm cut} = 0.1$ in the Lund diagrams. We see that a large portion of soft, wide-angle radiation is isolated from the hard, collinear core. The bottom right panel of Figure~\ref{fig:Lund_bkg} with quenched gluon jets shows the maximum density of soft, wide angle radiation. These soft jet constituents near the jet boundary are precisely the ones that contribute to the modification of the jet shape in \jw \cite{KunnawalkamElayavalli:2017hxo}. Within the framework of this diagrammatic representation, one can also examine the harder branch after the soft-drop procedure and Figure~\ref{fig:Lund_hard} shows the corresponding Lund diagram. We see that the vacuum structure starts to emerge in the hard branch of soft-drop PbPb jets. However, soft-drop still can not remove all the soft radiation within the hard branch in PbPb collisions. A similar observation was made in \cite{Hoang:2017kmk} that hard branchings can ``protect" soft radiation from being dropped. This is shown by the enhanced, wide angle radiation, even though now its angular scale has significantly reduced. The radiation can be due to correlated medium responses but is often removed with a statistical treatment of background in experiments.

The above studies demonstrate that telescoping deconstruction provides a physical organization of jet information where qualitatively new features can be captured order-by-order in the subjet expansion. It systematically improves quark gluon discrimination, leading to insights about jet quenching mechanisms as implemented in Monte Carlo simulations. Going beyond to higher T$N$ orders or using subjet superstructure \cite{Gallicchio:2010sw} and subjet charge \cite{Krohn:2012fg} information are beyond the scope of this work and left for future studies.

\section{Quark and Gluon Jet Modification and Discrimination}
\label{sec:results}

In the previous section we discussed jet representations using multiple, physics-motivated variables, jet images, telescoping deconstruction and Lund diagrams, and we showed how individual observables can be used to highlight the differences between quark jets and gluon jets, as well as between jets in proton-proton and heavy ion collisions. In this section we will use these information collectively and exhaustively in the study of jet modification and quark gluon discrimination.

For quark gluon discrimination, the performances are shown using Receiver Operating Characteristic (ROC) curves which plot the gluon jet efficiency or fake rate as a function of the quark jet efficiency or purity. Each ROC curve is labeled with the corresponding area under the curve (auc). Note that a smaller auc corresponds to a better performance, with larger values of quark purity and smaller values of gluon fake rate.

\begin{figure}[h]
	\centering
	\includegraphics[width=0.8\textwidth]{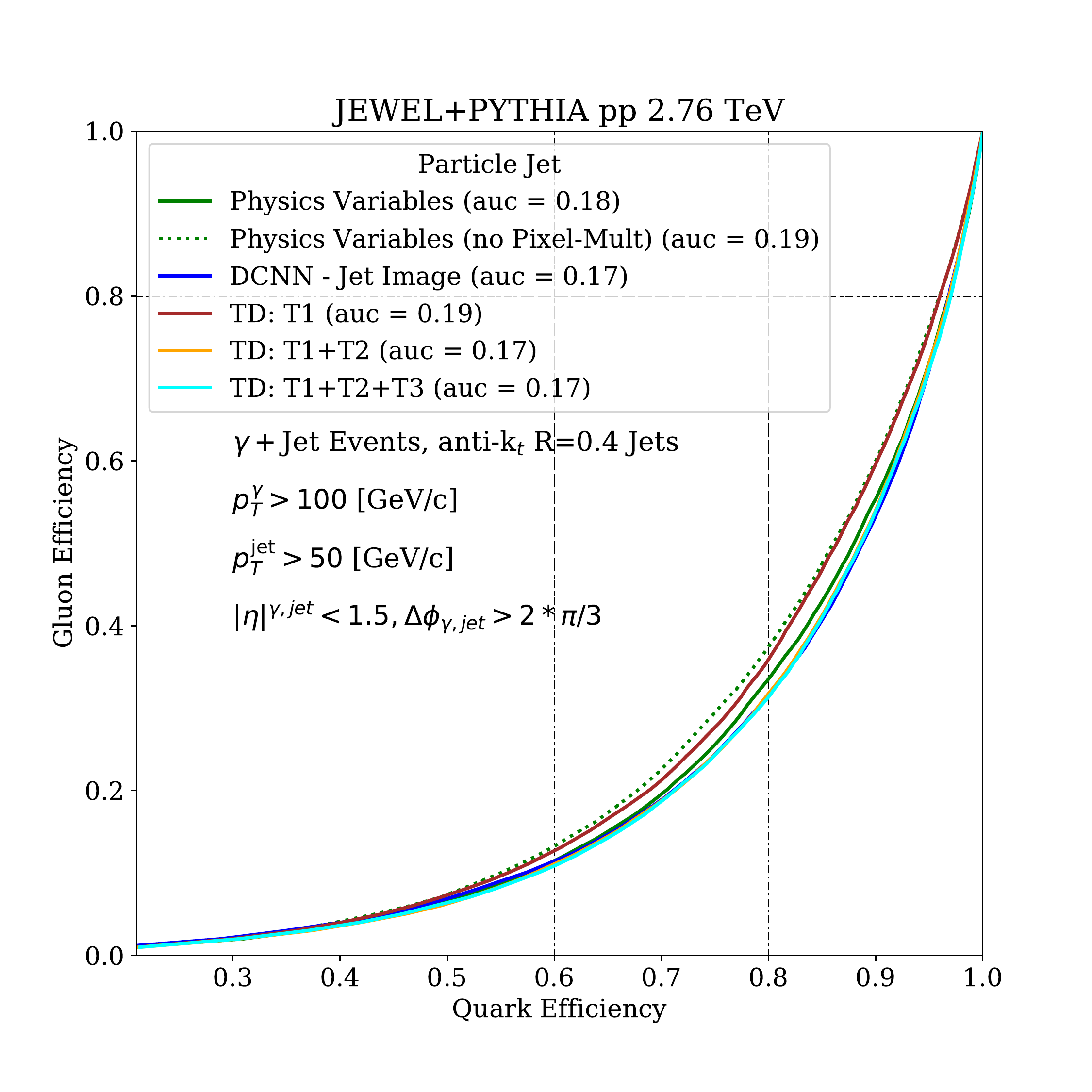}
	\caption{ROC curves of quark v.s. gluon efficiency for particle jets in pp collisions in \textsc{Jewel}. The different colored curves correspond to various classification models studied in this paper, and the quantity in the brackets marked by auc represents the area under the curve. The green solid and dotted lines are for the MLP with physics-motivated variables with and without the pixel multiplicity, respectively. The blue solid line corresponds to the DCNN trained on jet images, while the brown, orange and teal lines are for the TD at T1, T2 and T3 orders, respectively.}
\label{fig:ROC_pp}
\end{figure}

\begin{figure}[h]
	\centering
	\includegraphics[width=0.8\textwidth]{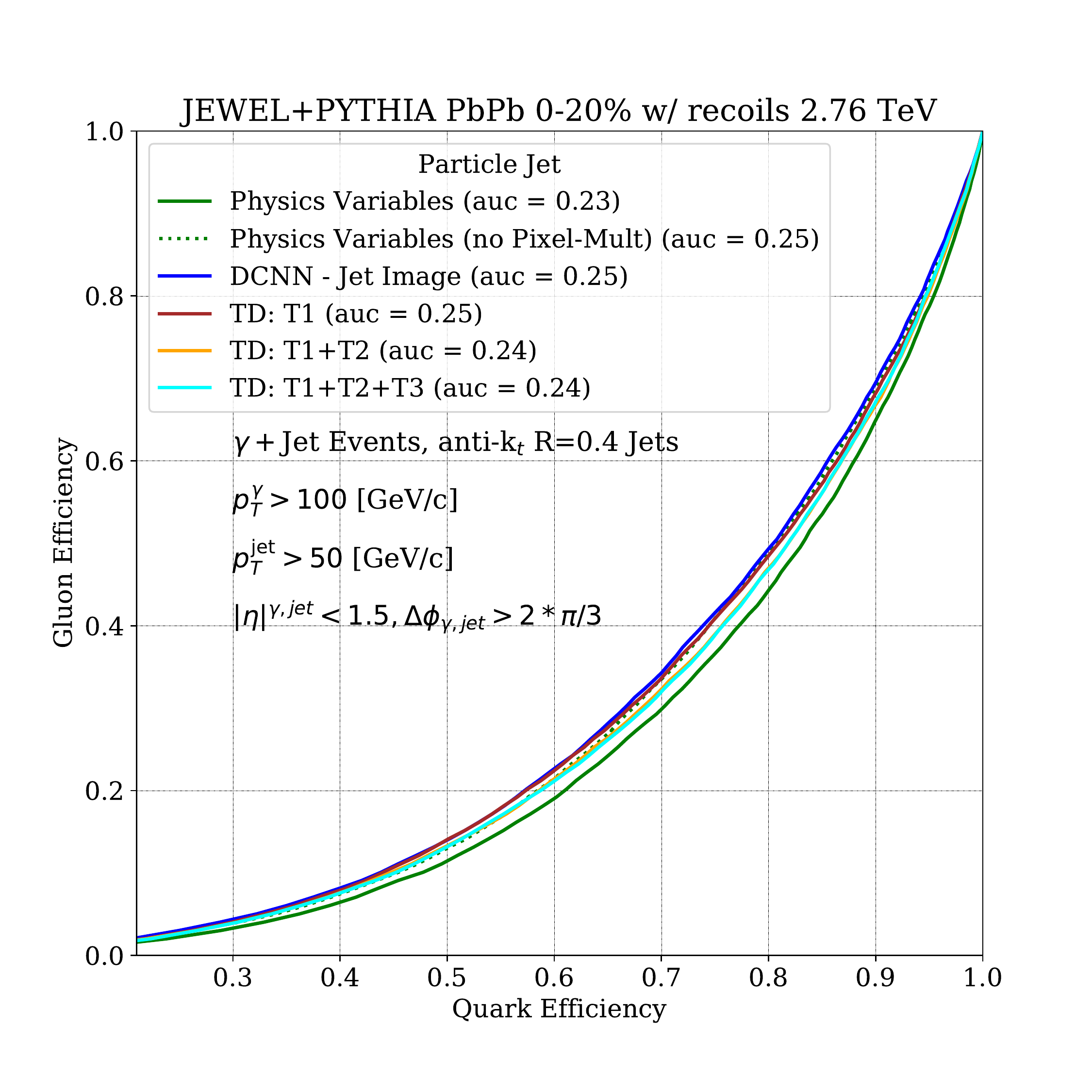}
	\caption{ROC curves of quark v.s. gluon efficiency for particle jets in central PbPb (right panel) collisions in \textsc{Jewel}, similar to Figure~\ref{fig:ROC_pp}.}
\label{fig:ROC_pbpb}
\end{figure}

Figure~\ref{fig:ROC_pp} and Figure~\ref{fig:ROC_pbpb} show ROC curves for quark gluon discrimination using physics-motivated variables, jet images and telescoping deconstruction in pp and PbPb collisions, respectively, for particle jets, i.e. jets reconstructed with particle momenta without $\eta-\phi$ discretization and background subtraction. We see a remarkable overall compatibility between the methods with similar classification performances, implying that the three jet representations capture comparable jet information. However, there are noticeable performance differences especially in their ordering. Note the monotonic increase of the telescoping deconstruction performances from $\rm T1$ to $\rm T1+T2$, to $\rm T1+T2+T3$ with increasing amount of input information of higher-order subleading subjets. It is important to note that the TD classification performance saturates quickly at the T2 order, with additional higher-order features offering little improvement in the quark efficiency given a gluon efficiency. In pp collisions (Figure~\ref{fig:ROC_pp}) DCNN outperforms the multivariate analysis using physics-motivated variables and has a similar performance as $\rm T1+T2+T3$. This implies that in the pp environment the $\eta-\phi$ discretization doesn't smear the relevant jet information, and the TD performance quickly saturates to the DCNN performance. In PbPb collisions (Figure~\ref{fig:ROC_pbpb}), the combination of physics-motivated variables gives better performance than telescoping deconstruction at the T3 order and DCNN due to the large particle multiplicity in the \jw PbPb events. Measuring the particle multiplicity in heavy ion collisions is highly nontrivial due to the huge underlying event contribution and its correlation to jets. To take into account the difficulty of such an observable, we also show the ROC curve (the dotted green line in Figure~\ref{fig:ROC_pp} and Figure~\ref{fig:ROC_pbpb}) for the MLP with physics-motivated variables excluding the pixel multiplicity, which reduces the auc by a few percent $0.01 - 0.02$. Even though the TD with radius scan in steps of 0.08 is not able to accurately capture the full multiplicity information at the T3 order, it is still within a few percent difference among other methods in the auc, providing confidence in its ability to capture the entirety of the jet's flavor dependent fragmentation.

\begin{figure}[h]
	\centering
        \includegraphics[width=0.8\textwidth]{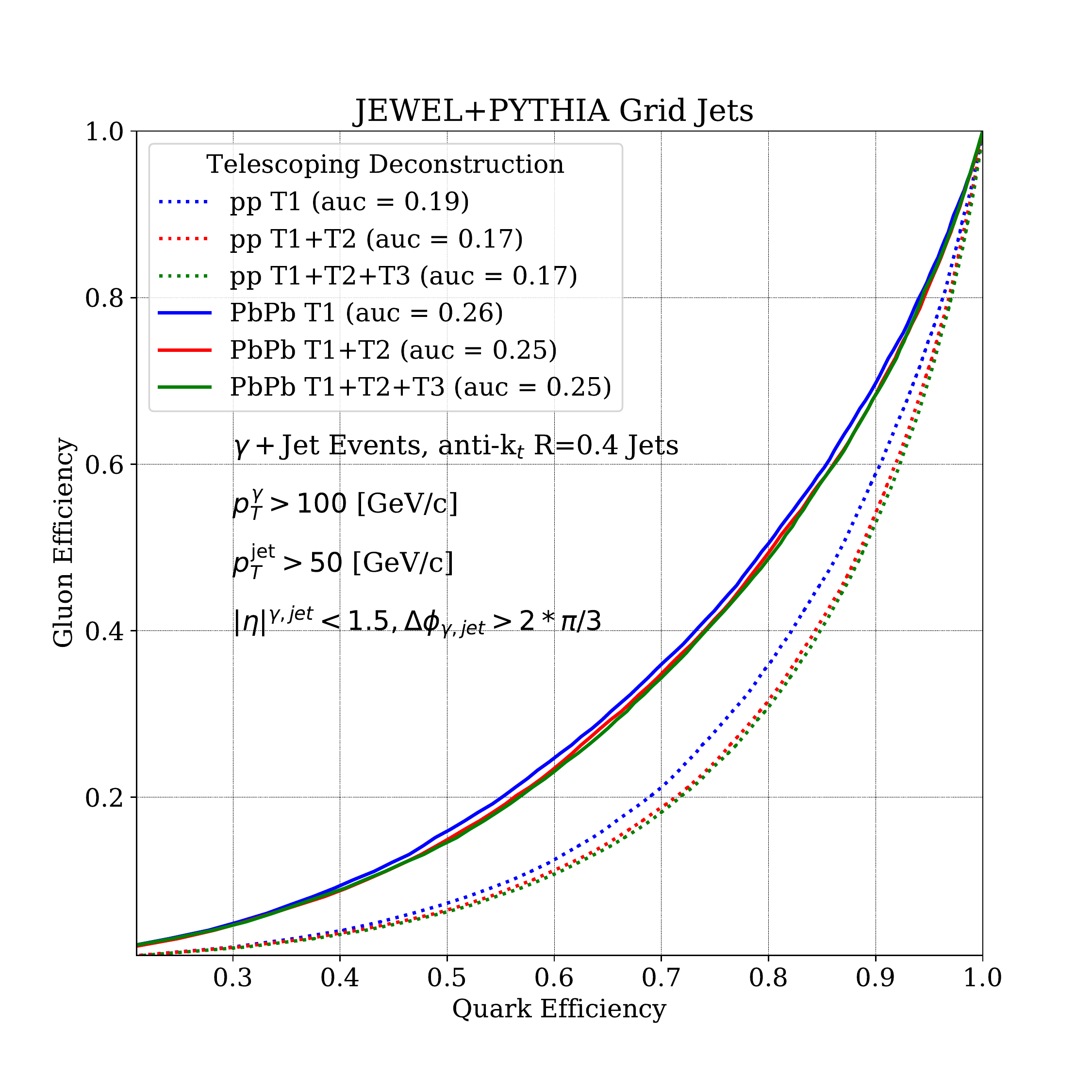}
	\caption{ROC curves of quark v.s. gluon efficiency using TD for grid jets in pp (dotted lines) and central PbPb (solid lines) collisions in \textsc{Jewel} with GridSub subtraction. The different colors represent different TD orders (T1: blue, T2: red, T3: green). The corresponding area under the curves are shown in the brackets.}
	\label{fig:ROC_TD_grid}
\end{figure}

\begin{figure}[h]
	\centering
        \includegraphics[width=0.8\textwidth]{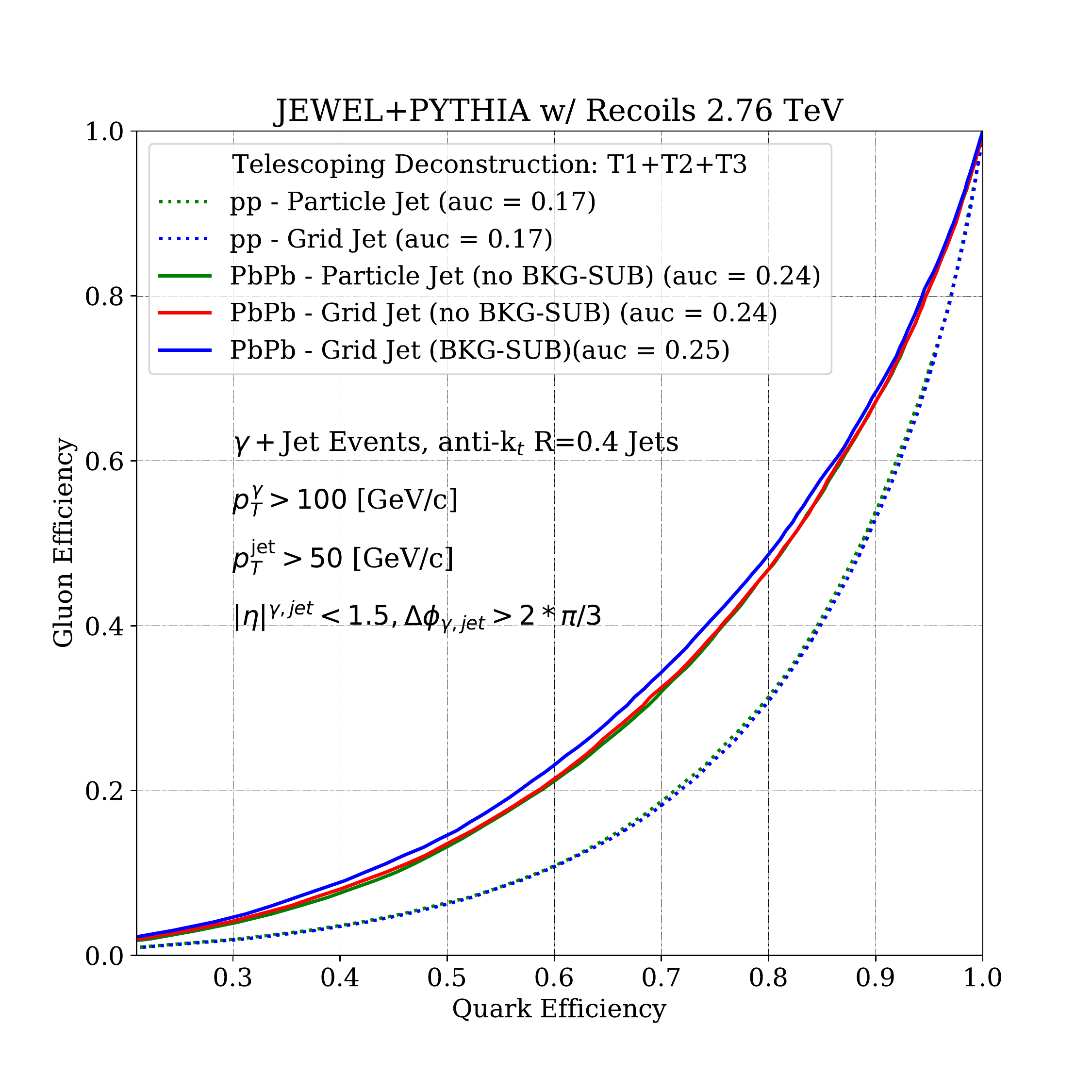}
	\caption{ROC curves of quark v.s. gluon efficiency using TD at the T3 order for jets with pp particle (green dotted), pp grid (blue dotted), PbPb particle (solid green), PbPb grid (solid red) and PbPb grid with GridSub subtraction (solid blue) constituents. The corresponding area under the curves are shown in the brackets.}
	\label{fig:ROC_TD3_grid}
\end{figure}

In order to have a fair comparison between jets in pp and PbPb collisions (and to directly compare with data), we apply the grid discretization for jets in both pp and PbPb collisions, the later with the specific GridSub background subtraction technique, and compare with various TD orders in Figure~\ref{fig:ROC_TD_grid}. We clearly see the performance drops in PbPb collisions, along with a significant reduction in the performance improvement from T1 to $\rm T1+T2$ in PbPb collisions. This suggests that the information carried in higher-order subleading subjets can be washed out in \jw heavy ion collisions due to the medium interactions. In Figure~\ref{fig:ROC_TD3_grid} we plot the ROC curves for TD at the T3 order and compare jets with or without $\eta-\phi$ discretization and background subtraction. Since the radius scan in TD respects the $\eta-\phi$ discretization resolution, we don't see the change of the TD performance using particle jets or grid jets. On the other hand, the performance drops very slightly (change in auc of $0.01$) with the background subtraction performed. This suggests that some intrinsic differences between quark jets and gluon jets are possibly removed in the subtraction~\footnote{Underlying events and their subtractions in experiments can both influence and introduce jet substructure features. Such effects can only be studied with full heavy ion event simulations, which we leave for future studies.}.

\begin{figure}[t]
	   \centering
	   \includegraphics[width=0.62\textwidth]{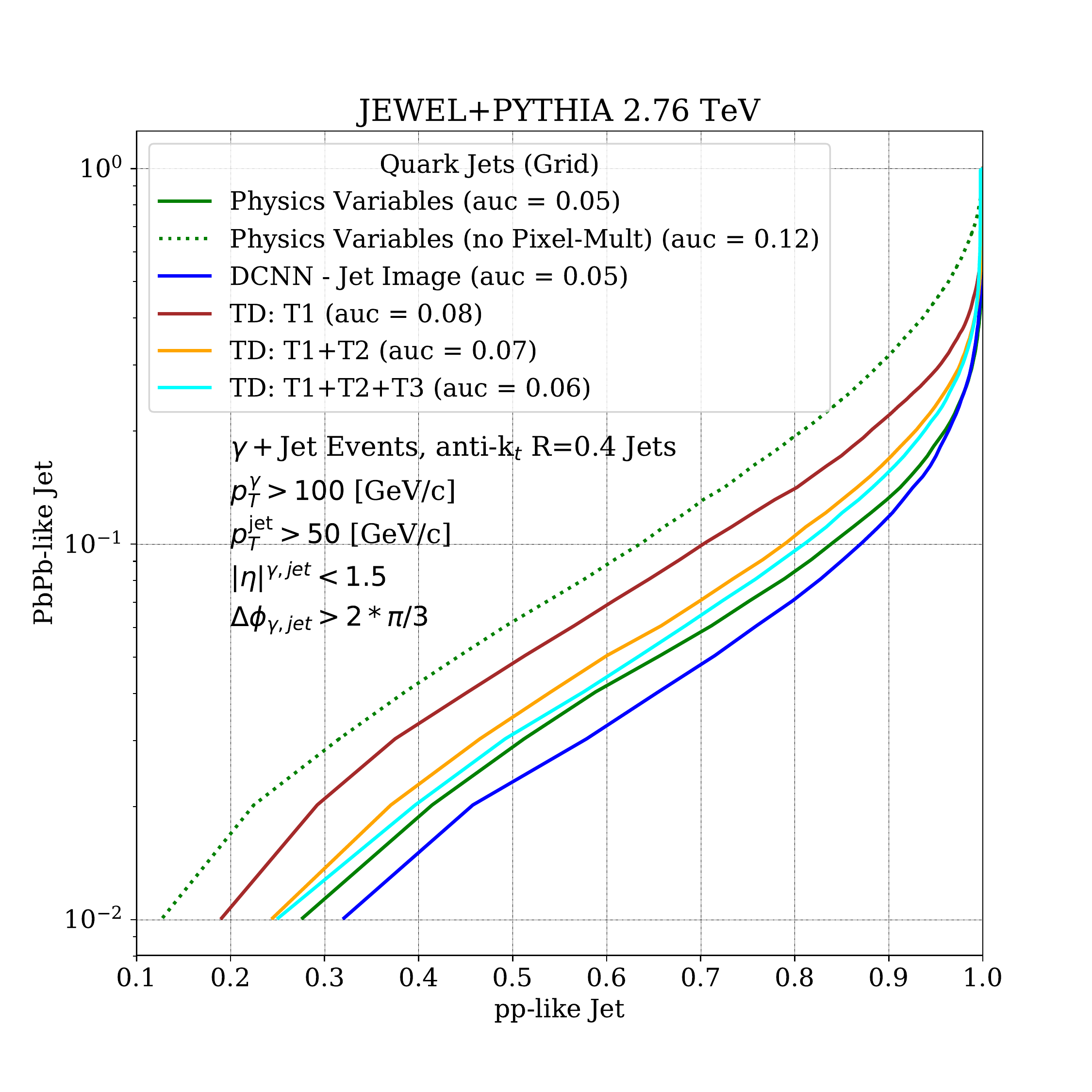}
	   \includegraphics[width=0.62\textwidth]{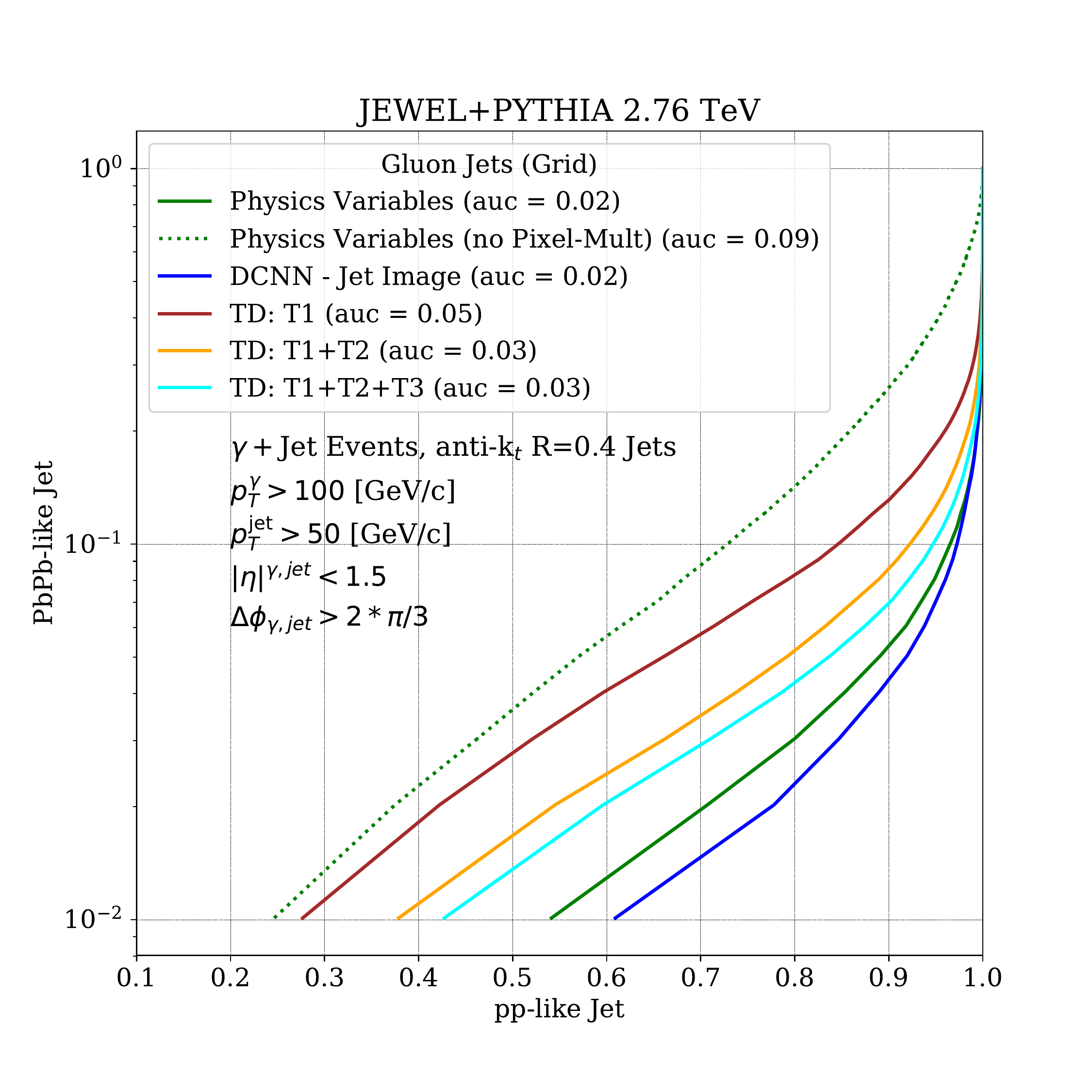}
	   \caption{ROC curves of pp-jet v.s. PbPb-jet efficiency for quark (top panel) and gluon (bottom panel) jets with grid discretization. The different colored curves correspond to various classification models studied in this paper.}
\label{fig:ROC_qq_gg}
\end{figure}

For a given jet flavor, i.e. quark or gluon jets, one can also study with the aforementioned methods, how the medium modifies jets by comparing vacuum jets with quenched jets in the context of proton-heavy ion jet discrimination. Figure~\ref{fig:ROC_qq_gg} shows the ROC curves plotting the PbPb jet efficiency as a function of the pp jet efficiency, for discriminating quark jets (top panel) or gluon jets (bottom panel) in pp and PbPb collisions using physics-motivated variables, jet images and telescoping deconstruction. We see huge differences between jets in pp and \jw PbPb collisions. All the methods perform significantly better in the task of identifying if a jet is quenched or not, compared to quark gluon discrimination. Therefore we plot the ROC curves using the log scale in the $y$-axis. We observe that the increase of the pixel multiplicity is a key feature in identifying quenched jets, with the DCNN performs slightly better than the MLP. The better performance in the proton-heavy ion jet discrimination for gluon jets (bottom panel of Figure~\ref{fig:ROC_qq_gg}) suggests that gluon jets are modified more significantly than quark jets due to medium-induced quenching.

From our comprehensive comparisons of quark v.s. gluon jets, as well as comparing vacuum v.s. quenched jets, we can qualitatively understand the effects of quenching as implemented in the \jw Monte Carlo. Quenched jets end up with significantly more jet constituents which are distributed away from the jet axis, resulting in the saturation of the TD classification performance at the T1 order. This modification appears to affect gluon jets significantly more than quark jets in \jw since a gluon jet, with its larger particle multiplicity and wider parton shower, can interact with a larger number of scattering centers leading to broader and more quenched jets.

\section{Conclusions}
\label{sec:conc}

We presented a systematic study of modifications to quark and gluon jet substructure in heavy ion collisions, taking the \jw Monte Carlo simulations as an example. Jet modifications are studied via the novel method of comparing and contrasting the classification of quark jets from gluon jets in different collision systems. We compare the performances of a MLP with physics-motivated jet variables, a state-of-the-art DCNN trained on jet images, and a newly developed method of telescoping deconstruction. We find that the quark gluon classification performance worsens for quenched jets, with a significant fraction of quark jets now masquerading as gluon jets. All the methods studied in this paper perform consistently, suggesting that the TD can extract fundamental jet fragmentation patterns. Through multiple methods and observables, we consistently find the dominant feature of the \jw jet quenching model to be the increase of soft particle multiplicity due to medium recoils throughout the jet region. This is closely related to the loss of information in subleading subjets, which is a characteristic feature of \textsc{Jewel}.

The future of heavy ion jet physics program necessarily moves toward precise, simultaneous understanding of multiple jet substructure observables and their correlations. Therefore our work serves as a first example of how multivariate techniques can help illuminate the underlying mechanism of jet modifications. By studying the telescoping deconstruction framework, we suggest new jet substructure observables for jet quenching studies, that we hope will provide qualitatively new insights order-by-order when measured experimentally. In the future, comparisons between theoretical calculations, simulations and experimental measurements will become standard practice in order to identify robust features of jet-medium interactions. Our work represents a thorough, systematic framework that we hope will serve this important purpose.

\section*{Acknowledgments}
The authors are grateful to Patrick Komiske and Eric Metodiev for collaborating on the development of telescoping deconstruction and contributions to the Appendix. The authors also thank Fr\'ed\'eric Dreyer, Benjamin Nachman, Joern Putschke, Jesse Thaler, Konrad Tywoniuk and Sevil Salur for helpful discussions. We thank Gavin Salam and the participants of the CERN TH institute ``{\sl Novel tools and observables for jet physics in heavy ion collisions}" and the 5-th Heavy Ion Jet Workshop for bringing to our attention the Lund Diagram and its usage in studying jet modifications. RKE thanks the organizers of the DS@HEP 2017 workshop at FermiLab for an excellent introduction on machine learning techniques. YTC also thanks the organizers of the BOOST2015 conference where the ideas of telescoping deconstruction and $\delta m$ were first presented. YTC was supported by the LHC Theory Initiative Postdoctoral Fellowship under the National Science Foundation grant PHY-1419008. RKE also acknowledges support in part from the National Science Foundation under Grant No.1067907 \& 1352081 and in part by the U.S. Department of Energy Office of Science, Office of Nuclear Physics under Award Number DE-FG02-92ER-40713.

\newpage
\appendix
\section*{Appendix}
\section{Probing subjet energy flows with telescoping deconstruction}

In this Appendix, we present the framework of telescoping deconstruction (TD) to systematically probe aspects of jet formation via its fragmentation basis. By scanning around dominant energy flows with multiple angular resolutions \cite{Chien:2013kca,Chien:2014hla}, one can efficiently quantify the radiation pattern. We demonstrated the use of the framework in the main text for quark/gluon discrimination and the study of medium modifications in heavy ion collisions. Here, to show the generality of TD we apply the method to different problems at the LHC: identifying and tagging boosted quark/gluon jets,  boosted $W$ and tops at the LHC. Crucially, the framework involves a fixed-order organization of individual observables which allows systematically improvable jet studies. Explicit examples of the $W$ isolation \cite{Chien:2017xrb} and exposing the $W$ boson in a top jet are presented. This highlights the physically meaningful nature of each TD observables and demonstrates their collective power as a representation.

Recently, there has been progress in utilizing the complete information in a jet in an unbiased way using advanced machine-learning techniques. Powerful multivariate approximators, such as neural networks, are capable of extracting useful features of the data relevant for a specific task. Examples include jet images with convolutional neural networks (CNNs)~\cite{Cogan:2014oua,deOliveira:2015xxd,Komiske:2016rsd,Kasieczka:2017nvn}, clustering histories with recurrent neural networks~\cite{Louppe:2017ipp}, complete sets of high-level observables with dense neural networks~\cite{Datta:2017rhs,Datta:2017lxt,Aguilar-Saavedra:2017rzt} and linear basis of energy flow polynomials \cite{Komiske:2017aww}. See \cite{Larkoski:2017jix} for a more complete summary and discussion of recent progress.

We aim to encapsulate relevant physics information in simple, physical observables derived with the TD framework. These variables follow the perturbative expansion of QCD and the parton shower picture and therefore encode perturbative and non-perturbative physics information. Similar to fixed-order perturbative expansions and parton shower splitting kernels \cite{Nagy:2017ggp}, the expansion is ordered by the number $N$ of exclusively reconstructed subjets. At each order, $N$ axes are determined by finding the dominant energy flow directions in the rapidity-azimuth plane~\cite{Stewart:2010tn,Chien:2013kca,Stewart:2015waa,Thaler:2015xaa}, which is partitioned into energy flow regions determined by the nearest axis. Jets at multiple angular resolutions are probed simply by the kinematics of subjets consisting of particles within different distances $R_T$ from the energy flow axes.

\begin{figure}[t]
\centering
\includegraphics[width=.32\columnwidth]{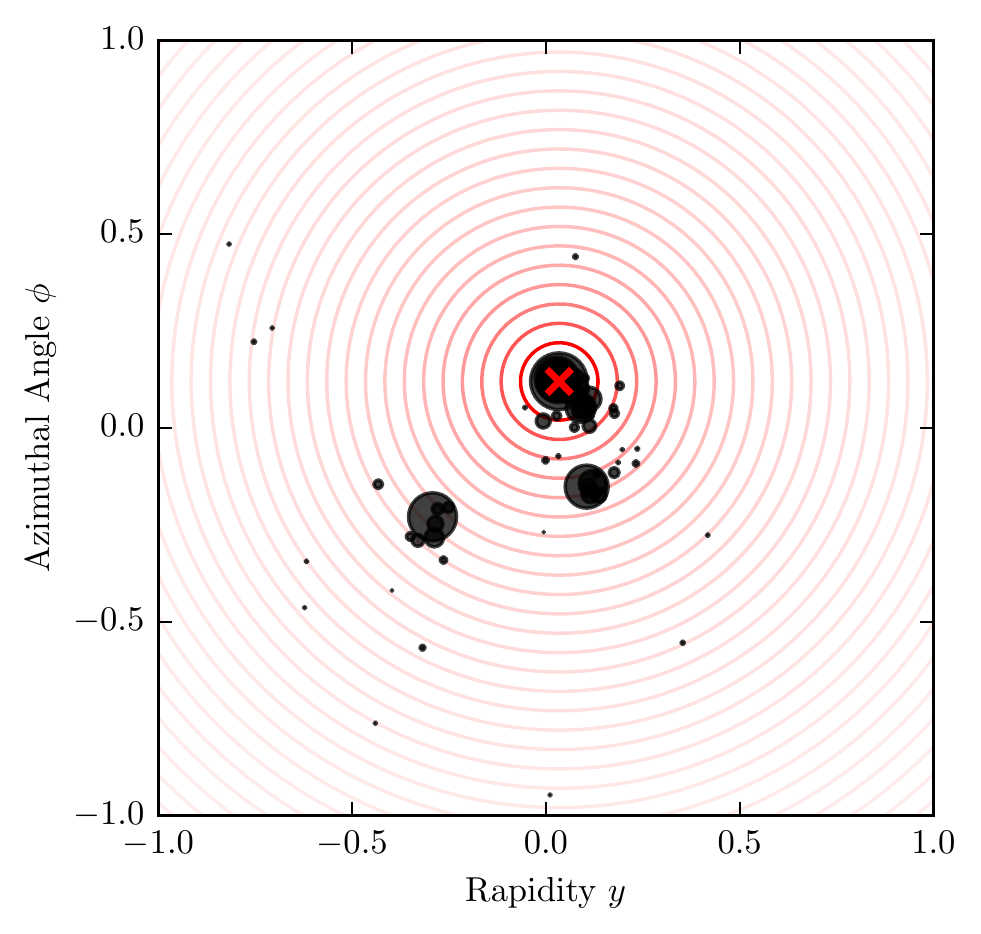}
\includegraphics[width=.32\columnwidth]{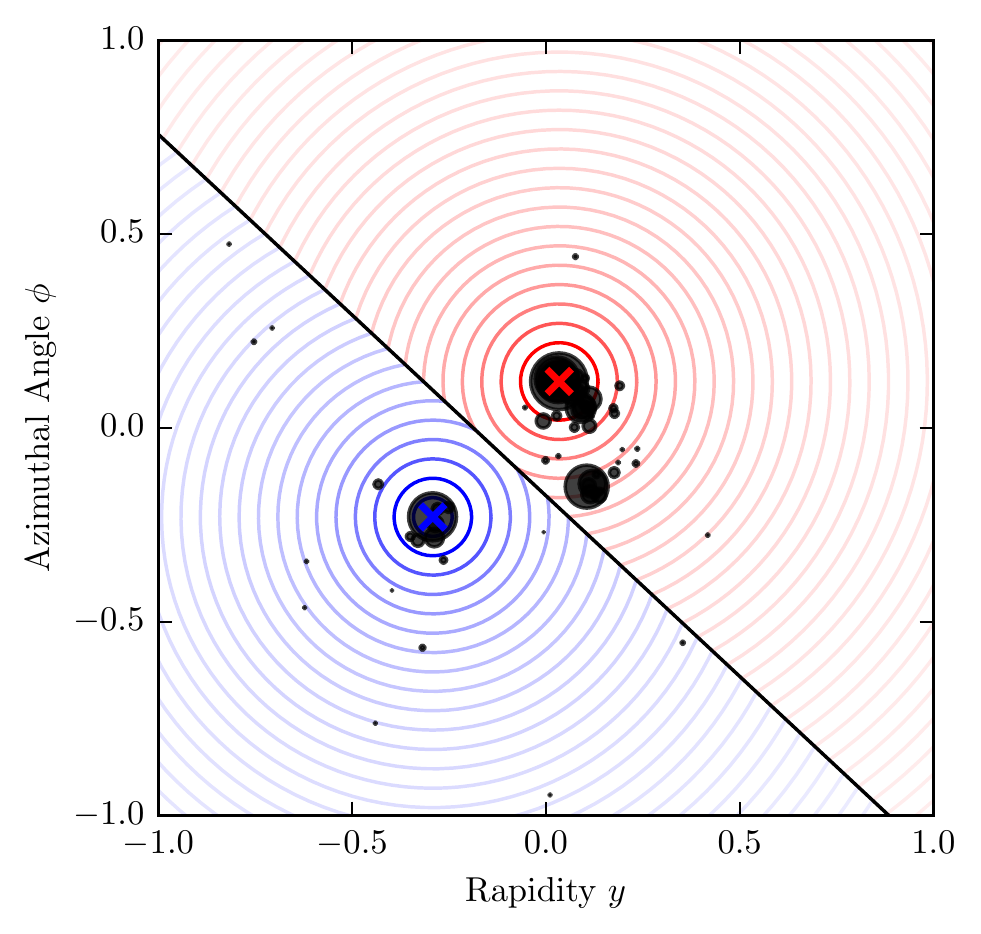}
\includegraphics[width=.32\columnwidth]{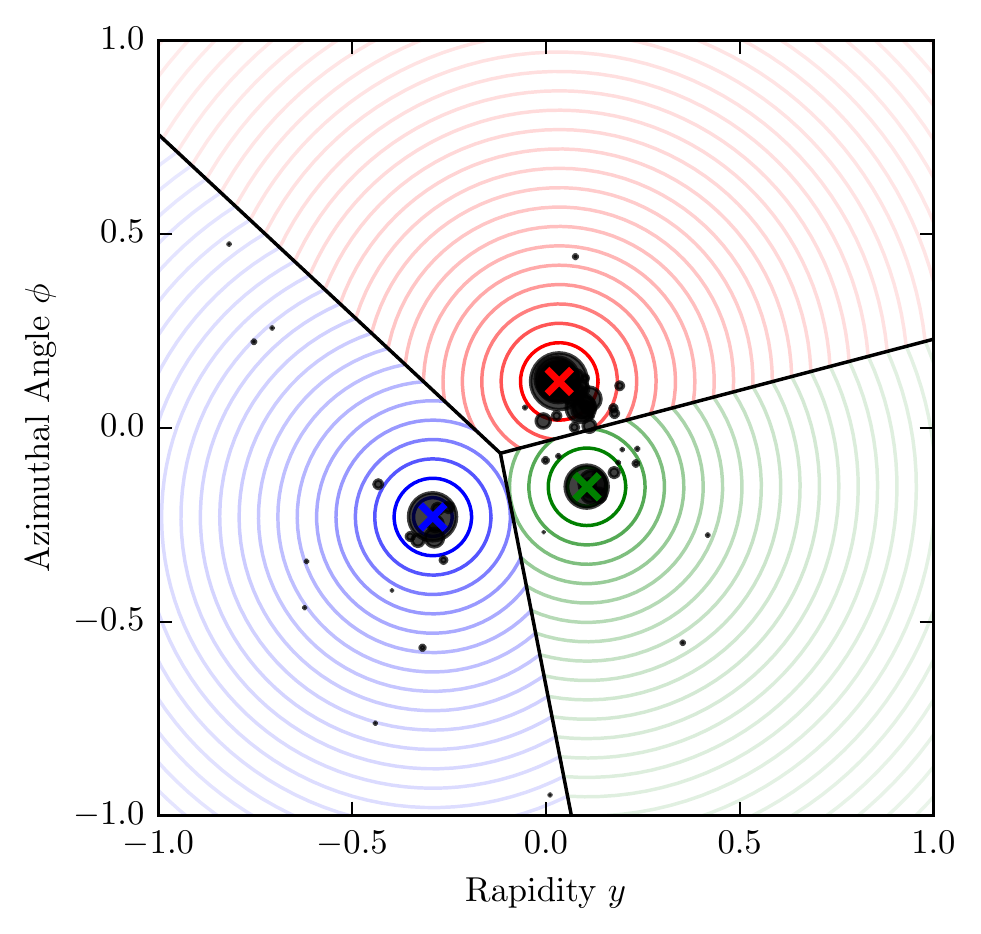}
\includegraphics[width=.32\columnwidth]{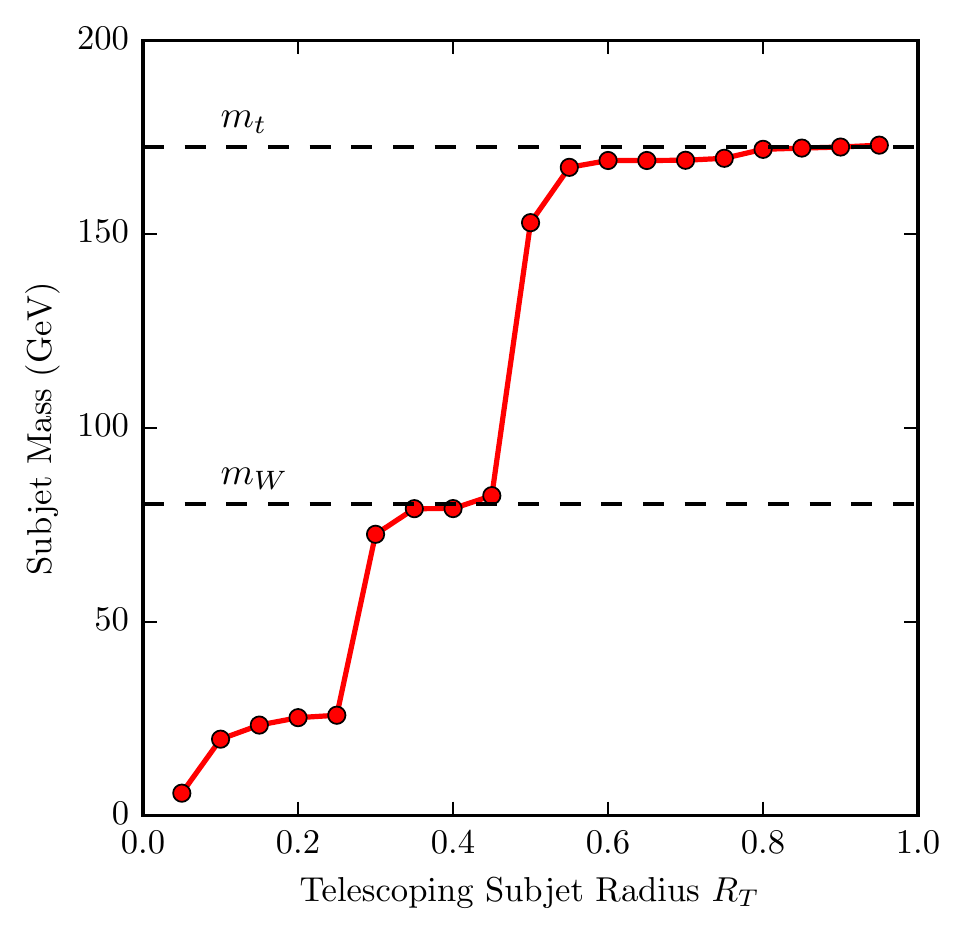}
\includegraphics[width=.32\columnwidth]{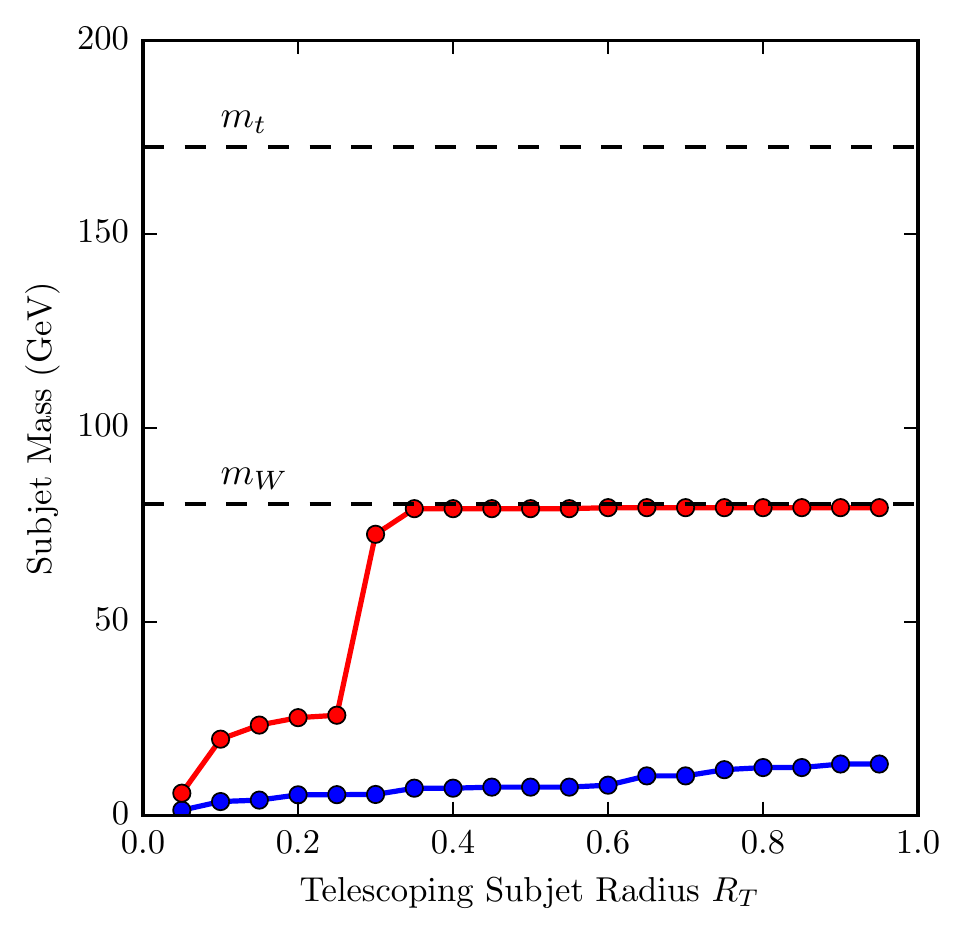}
\includegraphics[width=.32\columnwidth]{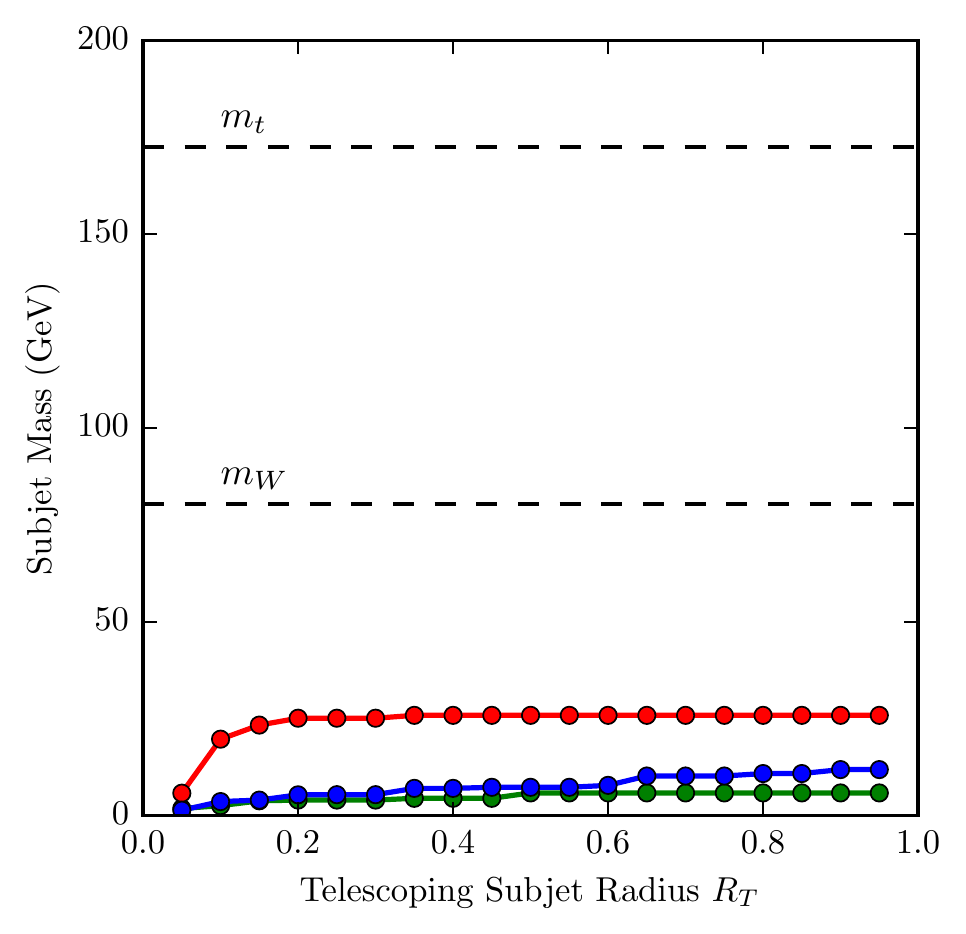}
\caption{\label{fig:tjet}(Top row) The TD of a top jet at T1 (left panel), T2 (middle panel) and T3 (right panel) orders. The crosses are the Winner-Take-All $k_t$ axes. The straight lines are the exclusive subjet boundaries. Particles are sized according to their transverse momenta. (Bottom row) The subjet masses at T1 (left panel), T2 (middle panel), and T3 (right panel) orders as the telescoping radius $R_T$ is varied, corresponding to the region of the same color in the top row. The presence of the $W$ in the top jet can be clearly seen at T1 and T2 orders by the plateau of the red line beginning around $R_T=0.3$.}
\end{figure}

The TD procedure is prescribed as follows:
\begin{itemize}
        \item At order $N$, determine $N$ axes $\{\hat n_i\}=\{(\eta_i,\phi_i)\}$ along the dominant energy flows, where $y$ and $\phi_i$ are the rapidity and the azimuthal angle of the subjet axis $i$. We use the Winner-Take-All (WTA) $k_T$ axes \cite{Bertolini:2013iqa}.
        \item Construct $N$ subjets with $M$ radii $\{R_{T,m}\}^M_{m=1}$ by assigning particles to the nearest axis according to the distance $d^2_{ij} = \Delta y_{ij}^2+\Delta \phi_{ij}^2$ between the axis $\hat n_i$ and the particle $j$ \cite{Stewart:2010tn,Chien:2013kca,Stewart:2015waa,Thaler:2015xaa}.
            \begin{equation}
                {\rm subjet}_{i,m} = \{p^\mu_j~|~d_{ij}<R_{T,m}~{\rm and}~d_{ij}<d_{kj}, \forall i\neq k\}.
            \end{equation}
            The subjet radii ${R_{T,m}}$ are sampled within the range $(0,R)$ where $R$ is the jet radius. In this paper $\{R_{T,m}\}$ are chosen to be evenly spaced within the range.
        \item We form the subjet data with the subjet transverse momenta and masses $\{(p_T,m)_{i,m}\}$,
            \begin{equation}
                {p_T}_{i,m}=\Big(\sum_{j\in~{\rm subjet}_{i,m}}p^\mu_j\Big)_T\;,~~~{m_{i,m}}^2=\Big(\sum_{j\in~{\rm subjet}_{i,m}}p^\mu_j\Big)^2\;,
            \end{equation}
            where we sum over all the particles $j$ within the subjet $i,m$. Together with the positions of the axes these form the telescoping deconstruction observables.
    \end{itemize}

The TD observables fall into two categories~\cite{Chien:2017xrb}: the {\sl subjet topology}, which is described by the axes and subjet transverse momenta, and the {\sl subjet substructure}, quantified by the subjet masses~\footnote{{\sl Subjet charge}~\cite{Krohn:2012fg} information can be included in this framework as well}. As the telescoping subjet order $N$ (T$N$ order) increases, more jet energy is covered by the subjets and the number of subjet radii sampled can be systematically decreased. In the large $N$ limit, the subjets reduce to individual particles with the full jet information. The telescoping deconstruction allows one to exploit features both within each subjet and among all the subjets. See Figure~\ref{fig:tjet} for the TD of a top jet for $N = 1$, $2$, and $3$ where the $W$ mass resonance can clearly be seen in the bottom panels.

To demonstrate the efficacy of TD in capturing the full jet information, we apply the framework in quark/gluon discrimination, boosted $W$ tagging and boosted top tagging. Each of these problems have signal jets with a different characteristic number of prongs: one, two, and three prongs, respectively. Events were generated from Monte Carlo simulation of proton-proton collisions at 14 TeV using \textsc{Pythia} 8.226~\cite{Sjostrand:2007gs}. Final-state non-neutrino particles are clustered into jets with \textsc{FastJet} 3.3.0~\cite{Cacciari:2011ma} using the anti-$k_t$ algorithm~\cite{Cacciari:2008gp}. We consider the boosted regime with the jet $p_T$ between 800 GeV and 900 GeV. For quark/gluon discrimination, quark jets were generated by $pp\to q+Z(\to\nu\bar\nu)$ and gluon jets by $pp\to g + Z(\to\nu\bar\nu)$, clustered into $R=0.4$ jets with rapidity $|y|<1.5$.

\begin{figure}[t]
\centering
\includegraphics[width=.32\columnwidth]{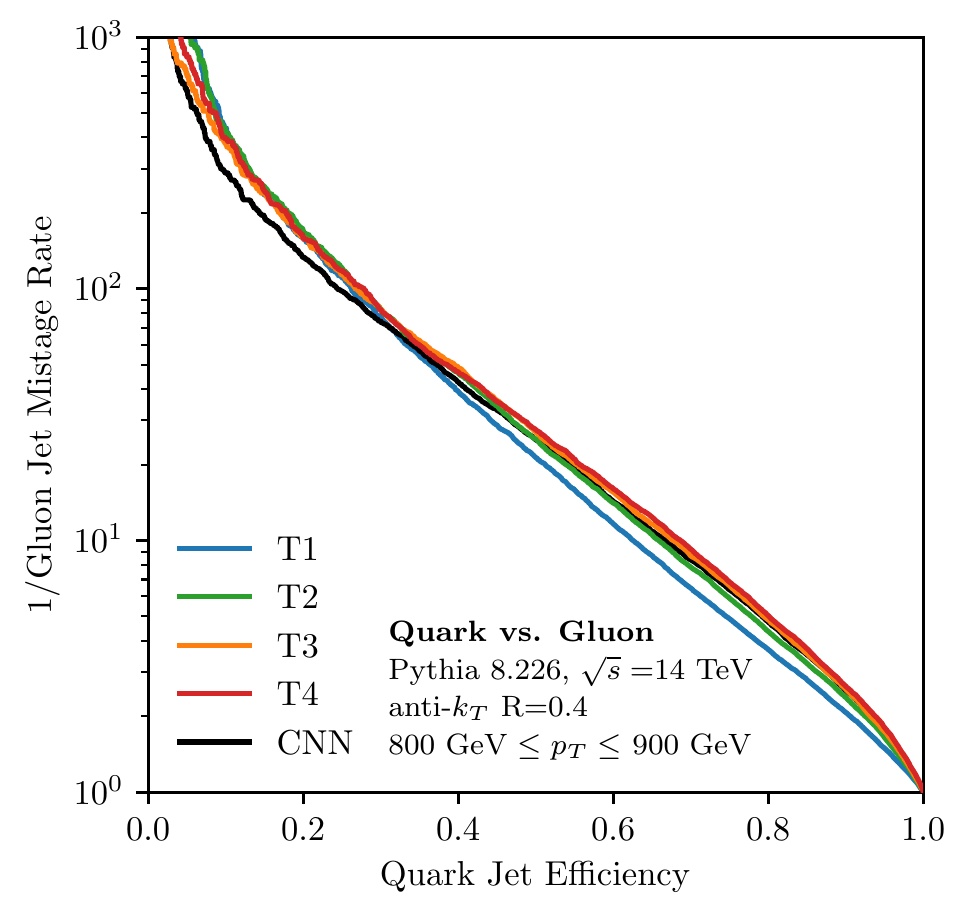}
\includegraphics[width=.32\columnwidth]{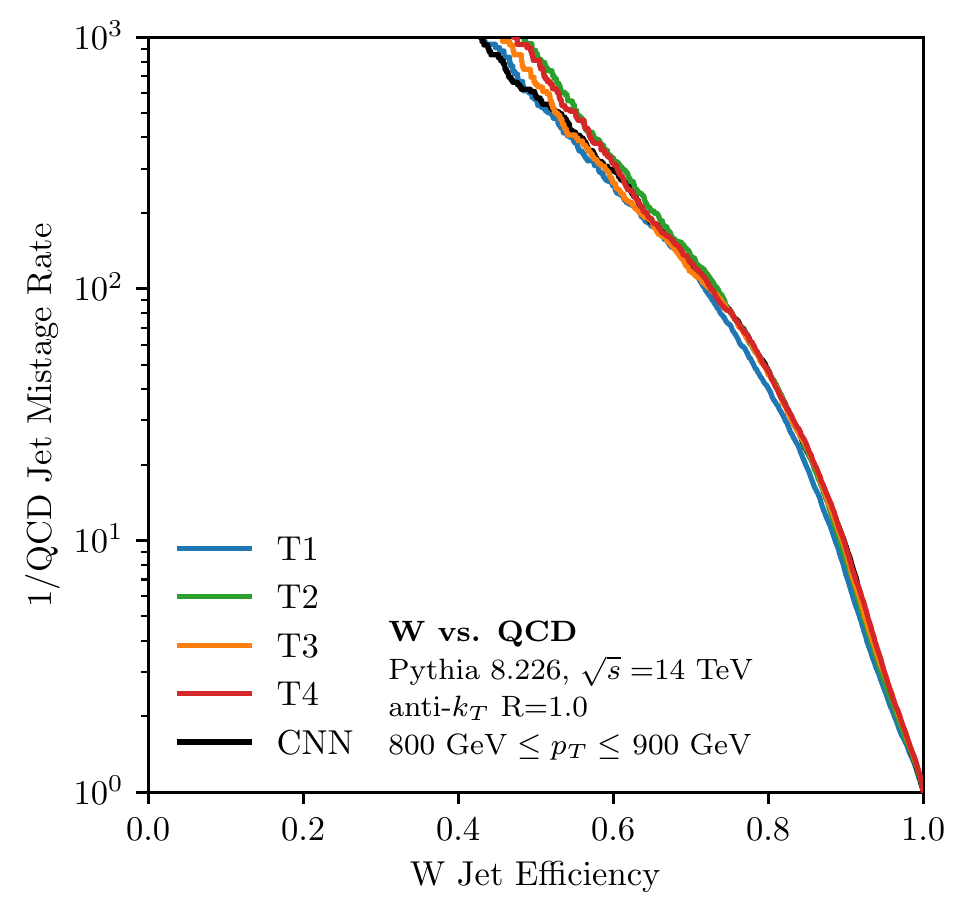}
\includegraphics[width=.32\columnwidth]{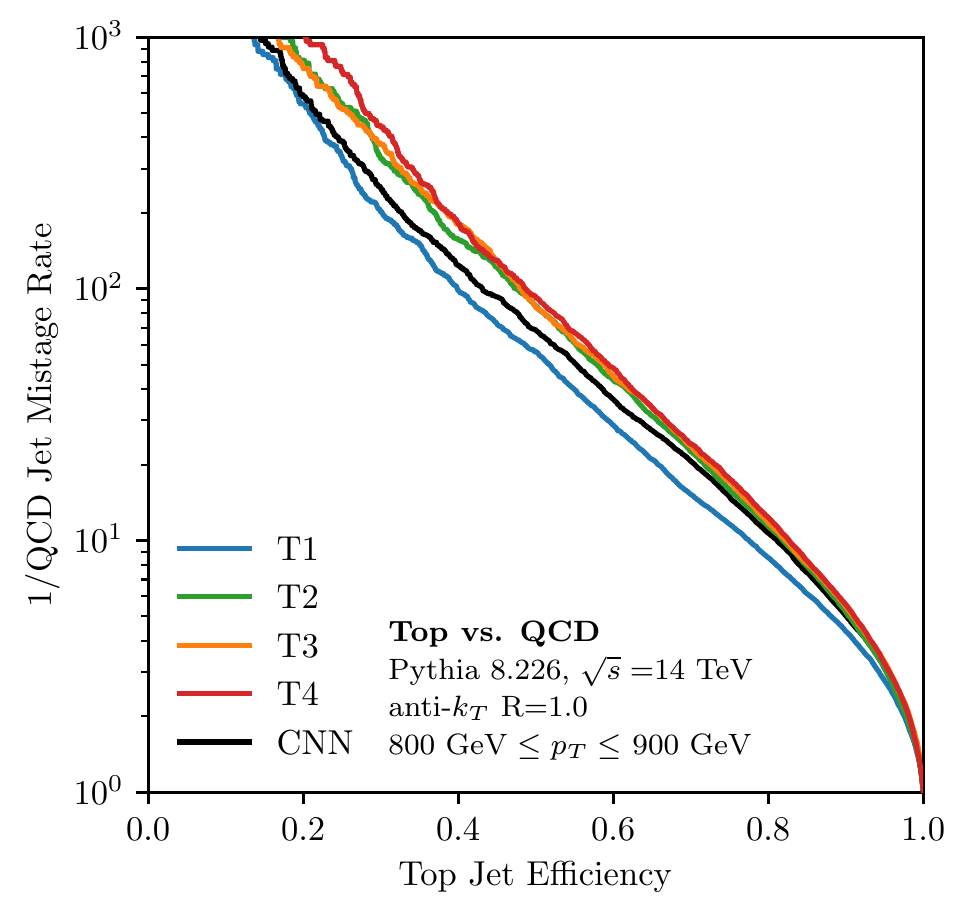}
\caption{\label{ROCs}ROC curves for the DNNs trained on cumulant telescoping deconstruction observables up to T1 through T4 orders and the jet image method using CNNs for quark/gluon discrimination (left panel), boosted $W$ (middle panel) and top (right panel) tagging. The T$N$ performance approximately saturates at T2 (quark/gluon), T1 ($W$ tagging), and T2 (top tagging) orders.}
\end{figure}

Since $W$ and top tagging are mass resonance searches where the jet mass is the most natural and powerful discriminating variable, we disentangle mass information to probe how much additional information can be exploited for tagging. Information from the hard process about the overall jet kinematics is eliminated by translating each jet to a frame where $y$ and $\phi$ of the jet are both zero. For $W$ and top tagging, $W$ jets were generated by $pp\to WW (\to \text{hadrons})$ and top jets were generated by $pp\to t\bar t (\to \text{hadrons})$, with background jets taken from QCD dijets. Final-state non-neutrino particles are clustered into $R=1.0$ jets with rapidity $|y|<1.5$. Signal and background jets identically populate a five-bin $1/p_T^4$ histogram in transverse momentum and a three-bin uniform histogram in mass between 75 GeV and 85 GeV for $W$ tagging and 160 GeV to 180 GeV for top tagging. Telescoping subjets are constructed with $R_T$ in steps of 0.05 between 0.05 and 0.4 in quark/gluon discrimination, and between 0.05 to 0.95 in $W$ and top tagging. 
For each problem, 200k events are generated for signal and background, with 10\% used for validation, 15\% for testing, and the remaining 75\% for training.

A neural network consisting of three dense layers with 100 nodes each is trained on the telescoping deconstruction observables up to T$N$ order. The training can be performed on a typical laptop CPU in fewer than five minutes. One could also use boosted decision trees (BDT) to combine variables \cite{Chien:2017xrb}. We compare the performance of our method to jet images using a CNN as done in the main text. The jet images are size $33\times 33$ and span a $2R\times 2R$ patch of the rapidity-azimuth plane with the intensity of each pixel corresponding to the total $p_T$ of particles in the pixel. The jet images are pre-processed and standardized according to the procedure in \Ref{Komiske:2016rsd}. The CNN architecture consists of three 48-filter convolutional layers with filter sizes of $8\times 8$, $4\times 4$, and $4\times 4$ followed by a 128-unit dense layer and a 2-unit softmaxed output layer. A $2\times 2$ maxpooling is performed after each convolutional layer with a stride length of 2. The dropout rate was taken to be 0.1 for all layers. All neural networks are implemented using the Python deep learning library Keras~\cite{keras} with the Theano backend~\cite{bergstra2010theano}. Rectified linear unit (ReLU) activation functions~\cite{nair2010rectified} and He-uniform model weight initialization~\cite{heuniform} are used. The networks are trained using the Adam algorithm~\cite{adam} with a learning rate of $10^{-3}$ and a batch size of 256 for 50 epochs with a patience parameter of 8, and the best model is selected based on validation set performance.

\begin{figure}[t]
\centering
\includegraphics[width=.32\columnwidth]{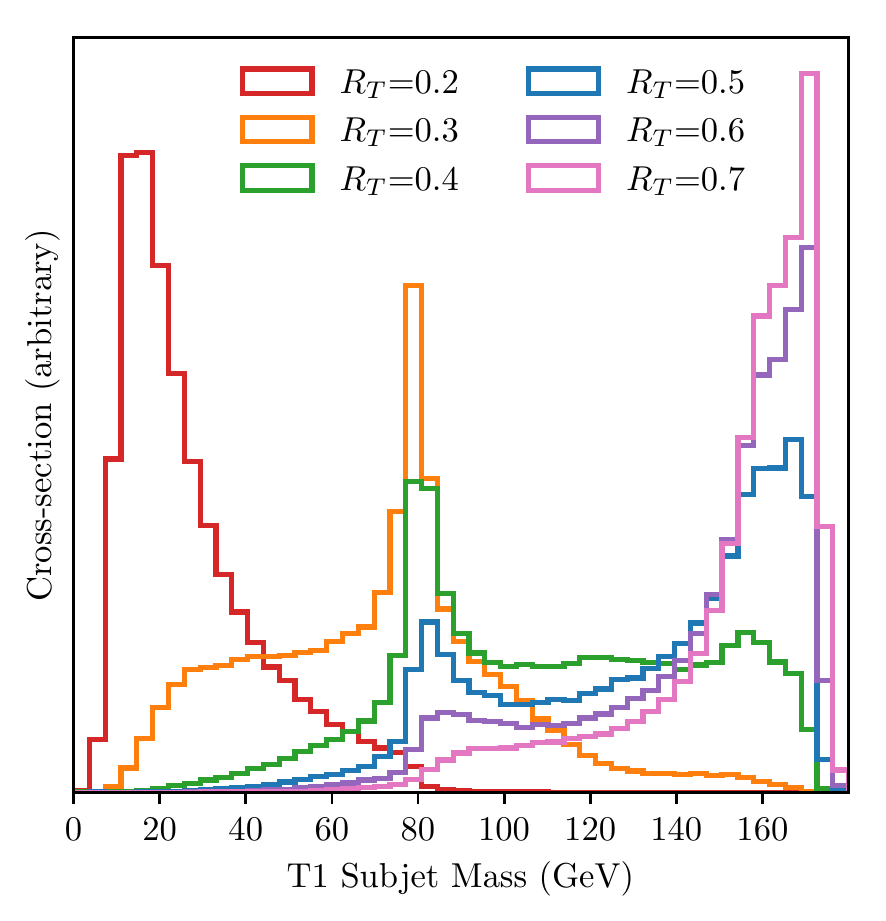}
\includegraphics[width=.32\columnwidth]{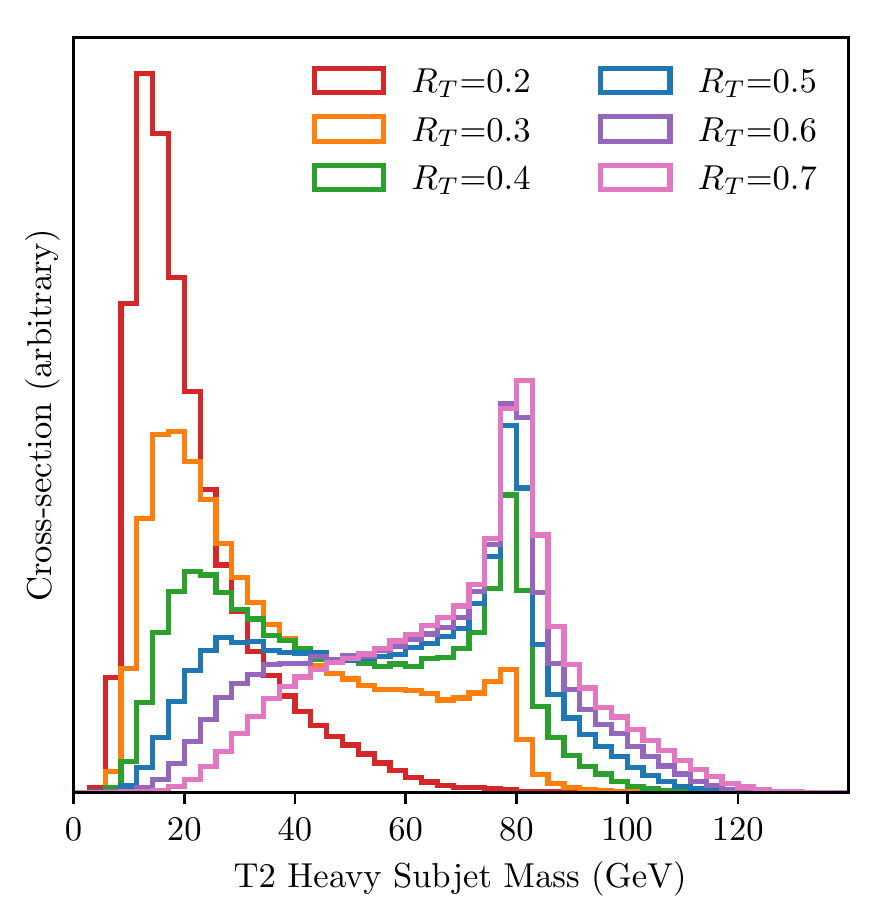}
\includegraphics[width=.32\columnwidth]{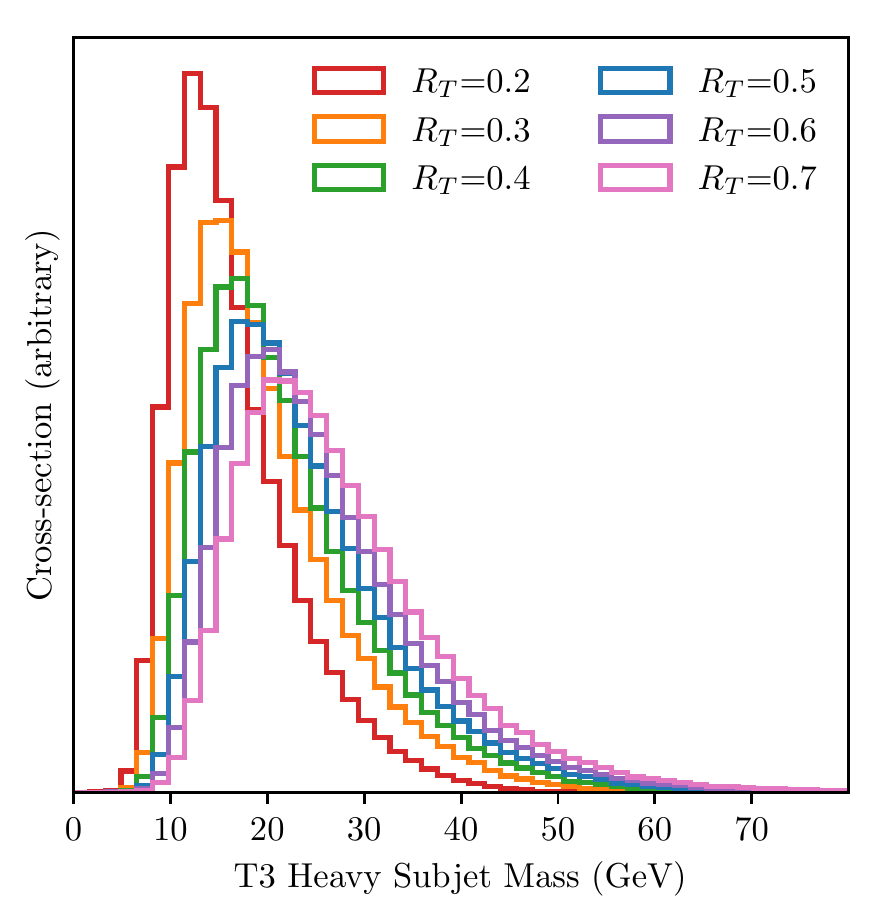}
\includegraphics[width=.32\columnwidth]{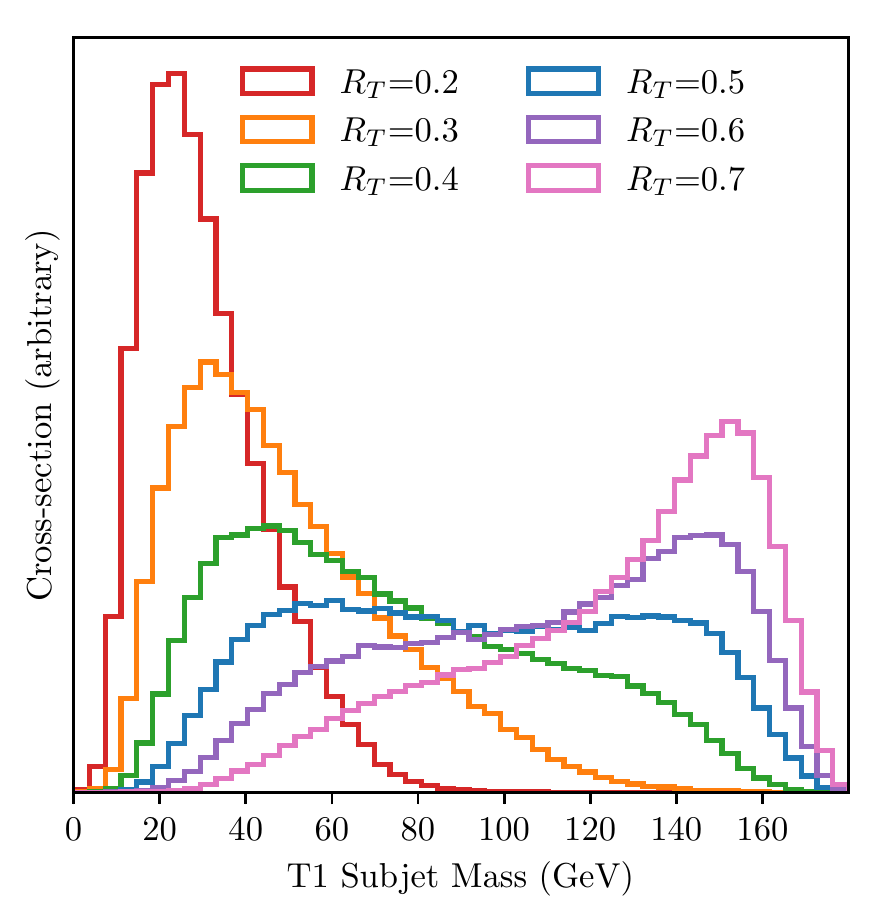}
\includegraphics[width=.32\columnwidth]{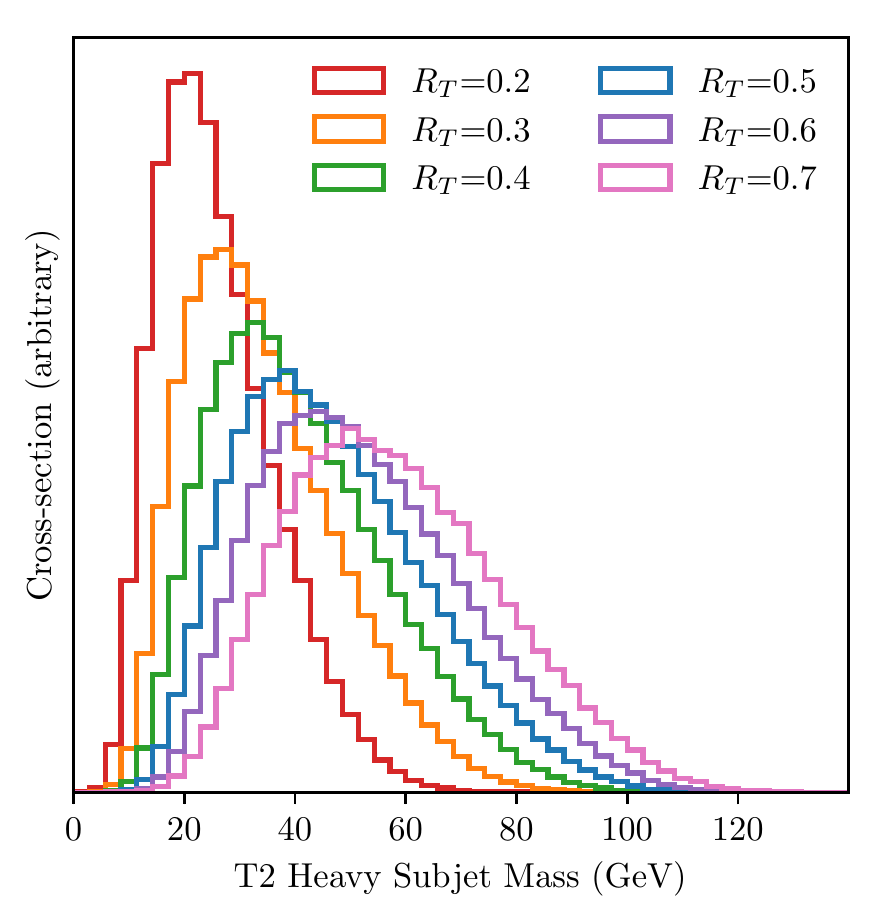}
\includegraphics[width=.32\columnwidth]{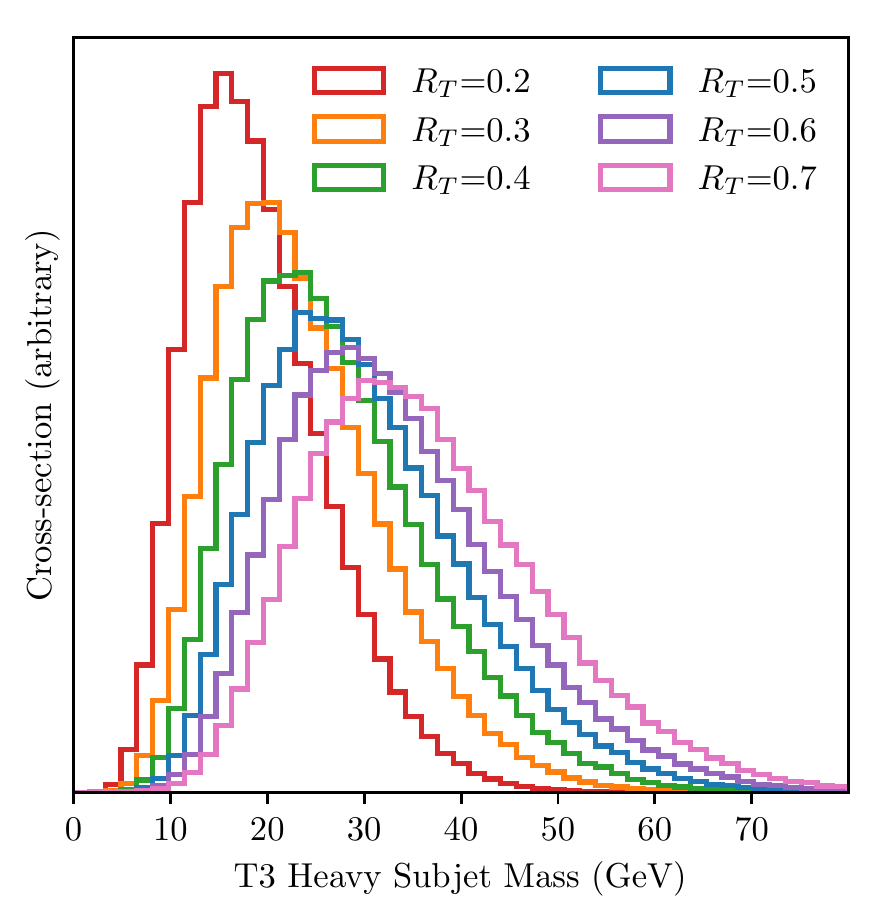}
\caption{\label{masses}The heavy subjet mass distributions at T1 (left panels), T2 (middle panels), and T3 (right panels) orders for multiple subjet radii $R_T$ of top jets (top row) and QCD background jets (bottom row). The presence of the $W$ in top jets is evident by the peaks at the $W$ mass at T1 and T2 orders. At T3 order, the heavy subjet mass distributions are QCD-like and are narrower (more quark-like) in top subjets than the wider (more gluon-like) QCD subjets. These features highlight the ability of the telescoping deconstruction to probe subjet substructure.}
\end{figure}

The performance of the trained models can be captured in a slightly different ROC curve (compared to the main text) where we plot the inverse of the background mistag rate at different signal efficiencies, thus a higher curve indicates better classification performance. Figure~\ref{ROCs} shows the ROC curves for the three tagging problems of the models trained on the cumulant telescoping deconstruction observables up to T$N$ order for $N\in\{1,2,3,4\}$ and the CNNs trained on jet images. The T$N$ performance converges quickly and is comparable to the performance of the jet images approach. The CNN architecture has not been tuned exhaustively, therefore its ROC curves serve to give a general sense of performance.

For top tagging, there is a significant increase in performance from T1 to T2 order in Figure~\ref{ROCs}, and the performance saturates beyond this order due to the sensitivity of T2 to the $W$ in top jets. For $W$ tagging, T1 order is sufficient to achieve most of the classification performance, which unambiguously confirms the $W$ isolation feature in the boosted regime compared to the QCD background~\cite{Chien:2017xrb}. Clearly, the T1 order probes the depletion of the radiation at large angles within $W$ jets, whereas the QCD background jets continue to acquire mass from radiation at large angles. For quark/gluon discrimination, there is a significant increase in performance from T1 to T2 and a smaller increase from T2 to T3 where the performance saturates, suggesting the usefulness of T2 subjet substructure and its sensitivity to subjet flavors. This confirms that the T$N$ expansion converges efficiently and TD faithfully represents the jet information.

In addition to being a useful jet representation, the TD allows physical information to be easily extracted. Figure~\ref{masses} shows the heavy subjet mass \cite{Chien:2010kc} distributions at T1, T2, and T3 orders, scanned over different telescoping radii $R_T$, for top jets and their QCD background. As $R_T$ is increased, the top jet T1 subjet mass distributions transition from QCD-like, to peaked at the $W$ mass, to peaked at the top mass. In contrast, the background QCD jets do not peak at the $W$ mass for any $R_T$ and transition from QCD-like to more top-like as they acquire mass at larger radii. The top jets at T2 order clearly and automatically show the $W$ peak which is completely absent in the background distributions. At T3 order, both top and background distributions appear QCD-like, with the wider background distributions due to the prevalence of gluon subjets. Thus we find that TD extracts relevant and significant subjet substructure information in order to efficiently perform at its task, i.e. quark jet v.s. gluon jet or boosted $W$/top jets v.s. QCD jets.

\bibliographystyle{JHEP3}
\bibliography{qg_ML_ref}

\end{document}